# Nearly Linear-Time Packing and Covering LP Solvers

(Achieving Width-Independence and $O(1/\varepsilon)$-Convergence)

Zeyuan Allen-Zhu
zeyuan@csail.mit.edu
Microsoft Research AI

Lorenzo Orecchia
orecchia@bu.edu
Boston University

November 4, 2014*

**Abstract**

Packing and covering linear programs (PC-LPs) form an important class of linear programs (LPs) across computer science, operations research, and optimization. In 1993, Luby and Nisan [25] constructed an iterative algorithm for approximately solving PC-LPs in *nearly linear time*, where the time complexity scales nearly linearly in $N$, the number of nonzero entries of the matrix, and polynomially in $\varepsilon$, the (multiplicative) approximation error. Unfortunately, existing nearly linear-time algorithms [2, 11, 24, 32, 36, 37] for solving PC-LPs require time at least proportional to $\varepsilon^{-2}$.

In this paper, we break this longstanding barrier by designing a packing solver that runs in time $\widetilde{O}(N\varepsilon^{-1})$ and covering LP solver that runs in time $\widetilde{O}(N\varepsilon^{-1.5})$. Our packing solver can be extended to run in time $\widetilde{O}(N\varepsilon^{-1})$ for a class of well-behaved covering programs. In a follow-up work, Wang *et al.* [35] showed that all covering LPs can be converted into well-behaved ones by a reduction that blows up the problem size only logarithmically.

At high level, these two algorithms can be described as linear couplings of several first-order descent steps. This is an application of our linear coupling technique (see [3]) to problems that are not amenable to blackbox applications known iterative algorithms in convex optimization.

*First version of this paper appeared on arXiv on this date. A 6-paged abstract of this paper, entitled "Nearly-Linear Time Positive LP Solver with Faster Convergence Rate," was presented at the STOC 2015 conference in Portland, OR. [1]. The journal version shall appear in Mathematical Programming.

# 1 Introduction

A packing linear program (LP) takes the form $\max\{c^T x : Ax \le b\}$ where $c \in \mathbb{R}^n_{\ge 0}$, $b \in \mathbb{R}^m_{\ge 0}$, and $A \in \mathbb{R}^{m \times n}_{\ge 0}$. A covering LP can be written as $\min\{b^T y : A^T y \ge c\}$, with the same requirements on $A, b$, and $c$. We denote by $N$ the number of non-zero elements in matrix $A$. We assume without loss of generality that the two LP programs are in their *standard forms*:

$$\text{Packing LP:} \qquad \max_{x \in \mathbb{R}^n_{\ge 0}} \{\mathbb{1}^T x : Ax \le \mathbb{1}\} \ , \tag{1.1}$$

$$\text{Covering LP:} \qquad \min_{y \in \mathbb{R}^m_{\ge 0}} \{\mathbb{1}^T y : A^T y \ge \mathbb{1}\} \ . \tag{1.2}$$

The two programs are dual to each other, so we denote by $\mathsf{OPT} \ge 0$ their shared optimum. We say $x$ is a $(1-\varepsilon)$-approximation for the packing LP if $Ax \le \mathbb{1}$ and $\mathbb{1}^T x \ge (1-\varepsilon)\mathsf{OPT}$, and $y$ a $(1+\varepsilon)$-approximation for the covering LP if $A^T y \ge \mathbb{1}$ and $\mathbb{1}^T y \le (1+\varepsilon)\mathsf{OPT}$.

In this paper, we study first-order iterative methods for solving packing and covering linear programs (PC-LPs) efficiently.[1] Of course, it is possible to adopt the Interior Point or Ellipsoid methods to obtain approximate solvers with a $\log(1/\varepsilon)$ dependence on the number of iterations. However, the computational cost of such algorithms is typically high, as each iteration requires solving a linear system, and thus is not suitable for large-scale applications.

To address this issue, researchers have developed *iterative approximate* PC-LP solvers that achieve a better dependence on the problem size (e.g., nearly linear in $N$) at the cost of having a $\mathsf{poly}(1/\varepsilon)$ dependence on the approximation parameter $\varepsilon$. Such iterative solvers have been widely applied in approximation algorithms (e.g., MinSetCover [25], MaxSet, MaxDiCut, Max-$k$-CSP [33], bipartite matching), probabilistic checkable proofs [33], zero-sum matrix games [30], scheduling [32], graph embedding [32], flow controls [11, 12], auction mechanisms [38], wireless sensor networks [15], and many other areas. In addition, techniques developed in this line of research have inspired important results on other fundamental algorithmic problems, such as the design of fast algorithms for multi-commodity flow problems [10, 19, 20, 26, 32] and the equivalence between QIP and PSPACE [22].

Previous iterative approximate solvers can be divided into two classes, *width-dependent* and *width-independent* solvers (see also Table 1).

**Width-Dependent Solvers[2] .** Based on multiplicative weight update ideas (a.k.a. exponentiated gradient updates), researchers have obtained solvers for PC-LPs with a running time at least $N$ multiplied with $\rho\mathsf{OPT} \in [1, \infty)$, where $\rho$ is the *width* of the program, i.e., the largest entry of matrix $A$. For instance, PC-LPs can be solved in $O(\frac{N\rho^2\mathsf{OPT}^2 \log m}{\varepsilon^2})$-time [32], or $O(\frac{N\rho\mathsf{OPT} \log m}{\varepsilon^2})$-time using some more refined analysis [8]. These algorithms only require "oracle-access" to the matrix $A$. When $A$ is given explicitly like in this paper, the running time can be reduced to $O(\frac{N\rho\mathsf{OPT} \log m}{\varepsilon})$ by deploying Nesterov's accelerated gradient method [30], or Nemirovski's mirror prox method [27].

Width-dependent algorithms are not polynomial time but only *pseudo-polynomial time.*

**Width-Independent, but Super Linear-Time Solvers.** Researchers also tried to appropriately scale the matrix so as to avoid the width penalty in the above methods. For instance, Bienstock and Iyengar [14] built on Nesterov's method [30] and obtained a running time $O(\varepsilon^{-1} N\sqrt{Kn \log m})$ where $K$ is the maximum number of non-zeros per row of $A$. This is $O(\varepsilon^{-1} N n \sqrt{\log m})$ in the worst case. The results of [16, 28] improved this complexity (for packing LP only) to $\widetilde{O}(\varepsilon^{-1} N \sqrt{n})$, at a cost of enduring an $\widetilde{O}(Nn)$-time preprocessing stage.

[1] Luby and Nisan, who originally studied iterative solvers for this class of problems [25], dubbed them *positive LPs*. However, the class of LPs with non-negative constraint matrices is slightly larger, including mixed-packing-and-covering LPs. For this reason, we prefer to stick to the PC-LP terminology.

[2] Most width-dependent solvers study the minmax problem $\min_{x \ge 0,\, \mathbb{1}^T x = 1} \ \max_{y \ge 0,\, \mathbb{1}^T y = 1} \ y^T A x$ , whose optimal value equals $1/\mathsf{OPT}$. Their approximation guarantees are often written in terms of *additive* error. We have translated their performances to multiplicative error for a clear comparison.

| Paper | Running Time | Width Independent? | Nearly Linear-Time? |
|---|---|---|---|
| Plotkin et al. [32] | $O(N \times \frac{\rho^2 \mathsf{OPT}^2 \log m}{\varepsilon^2})$ | no | no |
| Arora et al. [8] | $O(N \times \frac{\rho \mathsf{OPT} \log m}{\varepsilon^2})$ | no | no |
| Nemirovski [27], Nesterov [30] | $O(N \times \frac{\rho \mathsf{OPT} \log m}{\varepsilon})$ | no | no |
| Bienstock and Iyengar [14] | $O(N \times \frac{\sqrt{Kn \log m}}{\varepsilon})$ | yes | no |
| Nesterov [28]: *packing LP only* | $\widetilde{O}(N \times (n + \frac{\sqrt{n}}{\varepsilon}))$ | yes | no |
| Chudak and Eleutério [16]: *packing LP only* | $\widetilde{O}(N \times (n + \frac{\sqrt{n}}{\varepsilon}))$ | yes | no |
| parallel solvers [2, 9, 11, 12, 25, 35, 36] | $O(N \times \frac{\log^2 N \log(1/\varepsilon)}{\varepsilon^2})$ at best | yes | yes |
| Young [36] | $O((md + N) \times \frac{\log N}{\varepsilon^2})$ [3] | yes | almost yes |
| Bartal et al. [11, 12] | $O(nm \times \frac{\log N}{\varepsilon^2})$ | yes | almost yes |
| Young [37] | $O(N \times \frac{\log N}{\varepsilon^2})$ | yes | yes |
| Koufogiannakis and Young [24] | $O(N + (n + m) \times \frac{\log N}{\varepsilon^2})$ | yes | yes |
| **Theorem 3.4** *packing LP* | $O(N \times \frac{\log N \log \varepsilon^{-1}}{\varepsilon})$ | yes | yes |
| **Theorem 5.3** *well-behaved covering LP* | $O(N \times \frac{\log N \log \varepsilon^{-1}}{\varepsilon})$ | yes | yes |
| **Theorem 6.6** *covering LP* | $O(N \times \frac{\log N \log \varepsilon^{-1}}{\varepsilon^{1.5}})$ | yes | yes |

Table 1: Comparisons among iterative approximate solvers for packing and covering LPs. The width $\rho \in [1/\mathsf{OPT}, \infty)$ is defined as the largest entry of the constraint matrix $A$.

**Width-Independent, Nearly Linear-Time Solvers.** Perhaps the most desirable complexity is a running time that is both independent of the width parameter $\rho$, and also nearly linearly scales with $N$. [4] This line of research was initiated by a seminal paper of Luby and Nisan [25], who gave an algorithm running in $O\big(\frac{N \log^2 N}{\varepsilon^4}\big)$ time with no dependence on the width $\rho$. This is also the first *nearly linear-time* approximate solver for PC-LPs, and also the first to run in parallel in nearly linear-work and polylogarithmic depth.

The parallel algorithm of Luby and Nisan was extended by [2, 9, 11, 34, 36]. Most notably, the algorithm of Wang *et al.* [34] runs in $O(\frac{\log^2 N \log(1/\varepsilon)}{\varepsilon^2})$ iterations, each costing a matrix-vector multiplication that can be implemented in $O(N)$ total work.

The ideas of Luby and Nisan also led to sequential width-independent, nearly linear-time PC-LP solvers [11, 12, 24, 36, 37]. Most notably, the algorithm of Koufogiannakis and Young [24] runs in time $O\big(N + \frac{\log N}{\varepsilon^2} \times (n + m)\big)$.

Despite the amount of work in this area, the $O(1/\varepsilon^2)$ convergence rate was established in 1997 [11, 12] and has not been improved since then. On a separate note, Klein and Young [23] showed that all Dantzig-Wolfe type algorithms have to suffer from a $O(1/\varepsilon^2)$ convergence rate. This lack of progress constitutes a significant limitation, as the $\varepsilon^{-2}$-dependence (also known as the $1/\sqrt{T}$ convergence) on the approximation parameter $\varepsilon$ is particularly poor.

[3] The parameter $d$ is the maximum number of constraints each variable is in; $md$ may be larger than $N$.

[4] Some of these solvers still have a $\mathsf{polylog}(\rho)$ dependence. Since each occurrence of $\log(\rho)$ can be replaced with $\log(nm)$ after slightly modifying the matrix $A$, we have done so in Table 1 for a fair comparisons.

## 1.1 Our Results

**Packing LP.** We present an algorithm `PacLPSolver` that runs in $O(\frac{\log(nm/\varepsilon)\log(1/\varepsilon)}{\varepsilon}N)$ total time. This gives the first width-independent, and the first nearly linear-time solver for packing LP with an $\varepsilon^{-1}$ convergence (i.e., an $1/T$ convergence). In contrast, no nearly linear-time algorithm has achieved any convergence rate faster than $\varepsilon^{-2}$ before our work.

Interestingly, the maximum (weighted) bipartite matching is just one instance of a packing LP. As a consequence, our `PacLPSolver` algorithm finds an approximate maximum bipartite matching in time $\widetilde{O}(m\varepsilon^{-1})$. This new matching algorithm, which arises purely from convex-optimization arguments, matches the running time of the best known combinatorial algorithm for maximum weighted bipartite matching [17].

**Covering LP.** A symmetric design of `PacLPSolver` gives rise to an algorithm $\texttt{CovLPSolver}^{\mathsf{wb}}$ with the same running time $O(\frac{\log(nm/\varepsilon)\log(1/\varepsilon)}{\varepsilon}N)$, but only solving *well-behaved* covering LP instances. At a high level, we say an instance is well-behaved if the constraint $A^T y \geq \mathbb{1}$ is "never redundant": for instance, if the optimal solution $y^*$ satisfies $C \cdot \mathbb{1} \geq A^T y^* \geq \mathbb{1}$ for some constant $C > 1$ then the covering LP is well-behaved. For the *general* covering LP *without* well-behavior assumptions, we propose a different algorithm `CovLPSolver` that runs in time $O(\frac{\log(nm/\varepsilon)\log(1/\varepsilon)}{\varepsilon^{1.5}}N)$. Again, we emphasize that no nearly linear-time covering LP solver can achieve a convergence rate faster than $\varepsilon^{-2}$ (or equivalently $O(1/\sqrt{T})$) before our work.

Remark. After the first version of this paper appeared on arXiv in 2014, Wang, Rao and Mahoney [35] showed all covering LPs can be converted into well-behaved ones, by blowing up the problem size logarithmically. In other words, they obtained a nearly linear-time covering LP solver with $\varepsilon^{-1}$ convergence by a reduction to $\texttt{CovLPSolver}^{\mathsf{wb}}$. Nevertheless, our `CovLPSolver`, being a direct method, may still be of practical and theoretical interests.

## 1.2 Main Challenge and Our Approach

**Width-Independence vs. Acceleration.** Previous solvers for PC-LPs are based on standard techniques in non-smooth optimization. They first implicitly or explicitly smoothen the objective, often by the entropy regularizer. Then, they minimize the resulting convex objective either via variations of full-gradient methods, yielding parallel algorithms, or via variations of coordinate-gradient methods, yielding sequential algorithms. The *main challenge* in previous work is to show that the width dependence can sometimes be completely removed for PC-LPs, if the underlying minimization method is designed cleverly.

Of course, the slower the convergence rate is, the easier it is to design nearly linear-time solvers. The $\varepsilon^{-4}$-convergence solver of Awerbuch and Khandekar [9] and the $\varepsilon^{-3}$-convergence solver of [2] are arguably the simplest nearly linear-time solvers at this point.

In this paper, we achieve the $\varepsilon^{-1}$ convergence that is typical for accelerated gradient descent over smoothened objectives [30], but without paying the width or any additional super-logarithmic factors. The challenge in this approach is to preserve the width-independence and the accelerated rate *at the same time*. We stress here that our algorithm is *not an instance* of any known variant of accelerated gradient descent[5]. Moreover, the incorporation of width-independence and Nesterov's acceleration requires significant effort, as witnessed by the lack of progress on this problem for the last 15 years.

Finally, our algorithms are not Dantzig-Wolfe type, so can overcome the $1/\varepsilon^2$ lower bound of Klein and Young [23].

**Our High-Level Approach.** Our approach is based on an improved convex formalization $f(x)$ of the PC-LP objective, together with our linear-coupling framework for designing efficient first-order methods [3] for minimizing $f(x)$.

The improved formalization shows that our smoothened objective $f(x)$ satisfies either the classical condition for Lipschitz smoothness or a different condition based on multiplicative change. This formalization also clarifies why width-independent algorithms exist in the first place. See Lemma 2.7 and the related discussion for more details.

The linear-coupling framework in our previous work [3] provides a different interpretation of Nesterov's acceleration for smooth optimization [29]. In a nutshell, this linear-coupling framework allows us to construct accelerated algorithms by coupling the executions of a gradient descent algorithm, yielding iterates $\{\mathsf{y}_k\}$ and a mirror descent step algorithm, with iterates $\{\mathsf{z}_k\}$. The name "linear coupling" stems from the fact that, at iteration $k+1$, the gradient of the objective is queried at a point $\mathsf{x}_{k+1}$, which is a linear combination of gradient and mirror steps, i.e., $\mathsf{x}_{k+1} = (1-\tau)\cdot z_k + (1-\tau)\cdot y_k$.

In this paper, we apply linear coupling in a very non-trivial manner. We design a gradient and a mirror descent step, each very specific to the underlying PC-LP problem. We also perform a coupling step $\mathsf{x}_{k+1} = (1-\tau)\cdot z_k + (1-\tau)\cdot y_k$, but need to design a different analysis to preserve width independence. None of these components has appeared in [3].

**Arithmetic Precision.** Throughout this paper, we assume exact arithmetic operations for presenting the cleanest proofs. If the updates are calculated within precision $\frac{1}{\mathsf{poly}(\varepsilon^{-1},n,m)}$, or equivalently when word size $O(\log(\varepsilon^{-1}+n+m))$ is used, our results still hold.[6]

**Roadmap.** We relax the packing LP in Section 2, and provide our packing LP solver in Section 3. We relax the covering LP in Section 4, and provide our covering LP solver in the well-behaved case in Section 5. In Section 6, we provide our full covering LP solver.

# 2 Relaxation of the Packing Linear Program

To solve packing LP, we minimize a relaxed version of the original LP, where the hard constraint $Ax \le 1$ is regularized by entropy and replaced by an exponential penalty function.

**Notations.** Recall that the packing LP in its standard form is $\max_{x\ge 0}\{\mathbb{1}^T x : Ax \le \mathbb{1}\}$. Let us denote by $\mathsf{OPT}$ the optimal value of this linear program, and $x^*$ any optimal solution. We say that $x$ is a $(1-\varepsilon)$-approximation for the packing LP if $Ax \le \mathbb{1}$ and $\mathbb{1}^T x \ge (1-\varepsilon)\mathsf{OPT}$.

[5] This can be verified by observing that our objective $f_\mu(x)$, to be introduced later, is not globally Lipschitz smooth, so that one cannot apply accelerated gradient descent directly.

[6] Due to space limitation, we quickly sketch why logarithmic word size suffices for our algorithms. On one hand, one can prove in an iteration, if $x$ is calculated with a small additive error $1/\mathsf{poly}(1/\varepsilon,n,m)$, then the objective $f(x)$ may increase only by $1/\mathsf{poly}(1/\varepsilon,n,m)$ in that iteration. The proof of this relies on the fact that (1) one can assume without loss of generality all entries of $A$ are no more than $\mathsf{poly}(1/\varepsilon,n,m)$ and (2) our algorithms ensure $f(x) < \mathsf{poly}(1/\varepsilon,n,m)$ for all iterations with high probability, so even though we are using the exponential functions, $f(x)$ will not change additively by much. On the other hand, one can similarly prove that each $\nabla_i f(x)$ can be calculated within an additive error $1/\mathsf{poly}(1/\varepsilon,n,m)$ in each iteration. They together imply that the total error incurred by arithmetic operations can be made negligible.

Throughout this paper, we use the indices $i \in [n]$ to denote the columns of $A$, and the indices $j \in [m]$ to denote the rows of $A$. We let $A_{:i}$ be the $i$-th column vector of $A$, and $A_{j:}$ the $j$-th row vector of $A$. Given any vector $x$, we denote by $\|x\|_A = \sqrt{\sum_{i\in[n]} x_i^2 \cdot \|A_{:i}\|_\infty}$ the $A$-norm of $x$. By simple scaling, we can assume without loss of generality that[7]

$$\min_{i\in[n]}\{\|A_{:i}\|_\infty\} = 1 \ . \tag{2.1}$$

We restrict the domain of $x$ and the range of $\mathsf{OPT}$ as follows.

**Fact 2.1.** *Define the bounding box* $\Delta_{\mathsf{box}} \stackrel{\text{def}}{=} \{x \in \mathbb{R}^n : x_i \in \big[0, \frac{1}{\|A_{:i}\|_\infty}\big]\}$. *Under assumption (2.1), we have* $\mathsf{OPT} \in [1, n]$ *and* $\{x : x \geq 0 \wedge Ax \leq \mathbb{1}\} \subseteq \Delta_{\mathsf{box}}$.

*Proof.* Suppose that $i^*$ is the column that achieves the smallest infinite norm $\|A_{:i}\|_\infty$ over all columns. Letting $x$ be such that $x_i = 1$ at $i = i^*$ and $x_i = 0$ at $i \neq i^*$, we claim that $x$ is a feasible solution for the packing LP (1.1), simply because $\|A_{:i^*}\|_\infty = 1$ according to (2.1). This feasible solution $x$ yields an objective value $\mathbb{1}^T x = 1$, proving that $\mathsf{OPT} \geq 1$. On the other hand, for any solution $x \geq 0$ satisfying $Ax \leq \mathbb{1}$, we must have $x_i \leq \frac{1}{\|A_{:i}\|_\infty}$ for each $i$. Therefore, $\mathbb{1}^T x \leq \sum_i \frac{1}{\|A_{:i}\|_\infty} \leq n$, proving that $\mathsf{OPT} \leq n$.

The inclusion $\{x : x \geq 0 \wedge Ax \leq \mathbb{1}\} \subseteq \Delta_{\mathsf{box}}$ is obvious, since the constraints $x \geq 0$ and $Ax \leq \mathbb{1}$ together imply $x_i \leq \frac{1}{\|A_{:i}\|_\infty}$ for every $i \in [n]$. □

This bounding-box constraint allows us to focus only on searching $x$ in $\Delta_{\mathsf{box}}$.

**Our Regularized Objective.** We now introduce the smoothed objective $f_\mu(x)$ that we minimize over $\Delta_{\mathsf{box}}$ in order to approximately solve packing LP. At a high level, this objective $f_\mu(x)$ turns each row of the hard, non-smooth LP constraint $Ax \leq \mathbb{1}$ into an exponential penalty function so that we only need to require $x \in \Delta_{\mathsf{box}}$ throughout the algorithm.

Formally, the packing LP can be written as the following minimization problem by introducing the Lagrangian variable $y \in \mathbb{R}^m$:

$$\min_{x\in\Delta_{\mathsf{box}}} \big\{ -\mathbb{1}^T x \ + \ \max_{y\geq 0}\{y^T Ax - \mathbb{1}^T y\}\big\} \ . \tag{2.2}$$

The problem can be now smoothened by introducing a concave regularizer over $y \geq 0$. We take this regularizer to be the generalized entropy $H(y) = -\sum_{j=1}^m y_j \log y_j + y_j$ over the first orthant $y \geq 0$, and minimize the following smoothened objective $f_\mu(x)$ over $x \in \Delta_{\mathsf{box}}$:

$$f_\mu(x) \stackrel{\text{def}}{=} -\mathbb{1}^T x \ + \ \max_{y\geq 0}\{y^T Ax - \mathbb{1}^T y + \boxed{\mu \cdot H(y)}\} \ . \tag{2.3}$$

Above, $\mu > 0$ is some smoothing parameter to be chosen later. By explicitly computing the maximization over $y \geq 0$, $f_\mu(x)$ can be rewritten as

**Fact 2.2.** $f_\mu(x) = \mu \sum_{j=1}^m e^{\frac{1}{\mu}((Ax)_j - 1)} - \mathbb{1}^T x$ .

We study the *minimization* problem on $f_\mu(x)$ over $x \in \Delta_{\mathsf{box}}$. Intuitively $f_\mu(x)$ captures the original packing LP (1.1) as follows. Firstly, since we want to maximize $\mathbb{1}^T x$, the negative term $-\mathbb{1}^T x$ shows up in $f_\mu(x)$. Secondly, if a packing constraint $j \in [m]$ is violated by $\varepsilon$, that is, $(Ax)_j \geq 1 + \varepsilon$, the exponential penalty in $f_\mu(x)$ introduces a penalty at least $\mu e^{\varepsilon/\mu}$; this will be a large penalty if $\mu \leq O(\varepsilon/\log(n/\varepsilon))$.

**Remark 2.3.** The use of exponential function at least traces back to [32] in 1991 (implicitly) and to [21] in 1994 (explicitly). The way most previous results minimize $f_\mu(x)$ is by taking a logarithm

[7]If $\min_{i\in[n]}\{\|A_{:i}\|_\infty\} = 0$ then the packing LP is unbounded so we are done. Otherwise, if $\min_{i\in[n]}\{\|A_{:i}\|_\infty\} = v > 0$ we scale all entries of $A$ by $1/v$, and scale $\mathsf{OPT}$ by $v$.

$g(x) = \log\big(\sum_{j=1}^m e^{((Ax)_j - 1)/\mu}\big)$, explicitly or implicitly arguing that $g(x)$ is Lipschitz smooth (i.e., $\|\nabla^2 f(x)\|$ is bounded), and then taking gradient descent.[8] Unfortunately, the Lipschitz smoothness parameter of $g(x)$ depends on the width of the LP, and thus first-order iterative approaches based on directly minimizing $g(x)$ are mostly width-dependent [8, 27, 30, 32]. One can also reduce the width parameter of $g(x)$ which yields super linear-time solvers [14, 16, 28].

In this paper, we directly perform gradient descent and mirror descent on $f_\mu(x)$ —without taking the logarithm. Note that traditional accelerated gradient methods [29, 30] should not be applied directly to minimize $f_\mu$ because it is not Lipschitz smooth.[9]

Our $f_\mu(x)$ incurs a regularization error. The next proposition bounds this error following a similar treatment in [2].

**Proposition 2.4.** *Let $\mu = \frac{\varepsilon}{4\log(nm/\varepsilon)}$ and recall $x^*$ is an optimal solution for packing LP.*

*(a) $f_\mu(u^*) \leq -(1-\varepsilon)\mathsf{OPT}$ for $u^* \stackrel{\text{def}}{=} (1-\varepsilon/2)x^* \in \Delta_{\mathsf{box}}$.*
*(b) $f_\mu(x) \geq -(1+\varepsilon)\mathsf{OPT}$ for every $x \in \Delta_{\mathsf{box}}$.*
*(c) If $x \in \Delta_{\mathsf{box}}$ satisfies $f_\mu(x) \leq -(1-\theta)\mathsf{OPT}$ for some $\theta \in [0,1]$, then $\frac{1}{1+\varepsilon}x$ is a $\frac{1-\theta}{1+\varepsilon}$-approximate solution to the packing LP.*

**Remark 2.5.** Our box constraint $x \in \Delta_{\mathsf{box}}$ is almost redundant for minimizing $f_\mu(x)$: whenever $x \geq 0$ and $f_\mu(x) \leq 0$, one should automatically have $x_i \leq \frac{1+\varepsilon}{\|A_{:i}\|_\infty}$. However, this constraint shall be used to make sure that our updates are always inside $\Delta_{\mathsf{box}}$.

*Proof of Proposition 2.4.*

(a) We have $\mathbb{1}^T u^* = (1-\varepsilon/2)\mathsf{OPT}$ by the definition of $\mathsf{OPT}$. from the feasibility $Ax^* \leq \mathbb{1}$ in the packing LP, we have $Au^* - \mathbb{1} \leq -\varepsilon/2 \cdot \mathbb{1}$, and can compute $f_\mu(u^*)$ as follows:
$$f_\mu(u^*) = \mu \sum_j e^{\frac{1}{\mu}((Au^*)_j - 1)} - \mathbb{1}^T u^* \leq \mu \sum_j e^{\frac{-\varepsilon/2}{\mu}} - (1-\varepsilon/2)\mathsf{OPT}$$
$$\leq \frac{\mu m}{(nm)^2} - (1-\varepsilon/2)\mathsf{OPT} \leq -(1-\varepsilon)\mathsf{OPT} \ .$$

(b) Suppose towards contradiction that $f_\mu(x) < -(1+\varepsilon)\mathsf{OPT}$. Since $f_\mu(x) > -\mathbb{1}^T x$, it must satisfy that $\mathbb{1}^T x > (1+\varepsilon)\mathsf{OPT}$. Suppose that $\mathbb{1}^T x = (1+v)\mathsf{OPT}$ for some $v > \varepsilon$. By the definition of $\mathsf{OPT}$, we must have that $Ax < (1+v)\mathbb{1}$ is broken, and therefore there exists some $j \in [m]$ satisfying that $(Ax)_j \geq 1+v$. In such a case, the objective
$$f_\mu(x) \geq \mu e^{v/\mu} - (1+v)\mathsf{OPT} = \frac{\varepsilon}{4\log(nm/\varepsilon)}\Big((\frac{nm}{\varepsilon})^4\Big)^{v/\varepsilon} - (1+v)\mathsf{OPT}$$
$$\geq \Big(\Big((\frac{nm}{\varepsilon}\Big)^2)^{v/\varepsilon} - (1+v)\Big)\mathsf{OPT} > 0$$
giving a contradiction to the assumption that $f_\mu(x) < 0$.

(c) Note that $x$ satisfies $f_\mu(x) \leq -(1-\delta)\mathsf{OPT} \leq 0$, and we first show $Ax \leq (1+\varepsilon)\mathbb{1}$. Let us assume that $v = \max_j((Ax)_j - 1) \geq 0$ because otherwise we will have $Ax \leq \mathbb{1}$. Under this definition, we have $Ax \le (1+v)\mathbb{1}$ and therefore $\mathbb{1}^T x \le (1+v)\mathsf{OPT}$ by the definition of $\mathsf{OPT}$. We compute $f_\mu(x)$ as follows.

[8] Note that some of the previous results (such as [8, 32]) appear to directly minimize $\sum_{j=1}^m e^{((Ax)_j-1)/\mu}$ as opposed to its logarithm $g(x)$. However, their per-iteration objective decrease is multiplicative, meaning it is essentially equivalent to performing a single gradient-descent step on $g(x)$ with additive objective decrease.

[9] The exact same $f_\mu(x)$ also appeared in our previous work [2], albeit without this smoothing interpretation and without the constraint $x \in \Delta_{\mathsf{box}}$. The techniques in [2] only leads to $\varepsilon^{-2}$ convergence (see Table 1).

$$f_\mu(x) \ge \mu e^{\frac{v}{\mu}} - (1+v)\mathsf{OPT} \ge \mu\Big((\frac{nm}{\varepsilon})^4\Big)^{v/\varepsilon} - (1+v)n = \frac{\varepsilon}{4\log(nm/\varepsilon)}\Big((\frac{nm}{\varepsilon})^4\Big)^{v/\varepsilon} - (1+v)n \enspace .$$

The above quantity is positive whenever $v \ge \varepsilon$, and therefore, to satisfy $f_\mu(x) \le 0$ we must have $v \le \varepsilon$, which is equivalent to $Ax \le (1+\varepsilon)\mathbb{1}$. Next, because $-\mathbb{1}^T x \le f_\mu(x) \le -(1-\delta)\mathsf{OPT}$, we know $\mathbb{1}^T x \ge (1-\delta)\mathsf{OPT}$. Letting $x' = \frac{1}{1+\varepsilon}x$, we both have that $x'$ is feasible (i.e., $Ax' \le \mathbb{1}$), and $x'$ has an objective $\mathbb{1}^T x'$ at least as large as $\frac{1-\delta}{1+\varepsilon}\mathsf{OPT}$. □

**Some Non-Standard Smoothness Properties.** The gradient and Hessian of $f_\mu(x)$ can be written in the following closed forms:

**Fact 2.6.** *$\nabla f_\mu(x) = A^T p(x) - \mathbb{1}$ and $\nabla^2 f_\mu(x) = \frac{1}{\mu}A^T \mathrm{diag}\{p(x)\}A$, where $p_j(x) \stackrel{\mathrm{def}}{=} e^{\frac{1}{\mu}((Ax)_j - 1)}$.*

By staring at these closed forms, we note that $f_\mu(x)$ is not Lipschitz-smooth: for instance, each $\nabla^2_{ii} f_\mu(x)$ can go to infinity so the spectral norm of $\nabla^2 f_\mu(x)$ is unbounded. However, the non-negativity of $A$ guarantees that whenever $\nabla^2_{ii} f_\mu(x)$ is large for some coordinate $i$, the corresponding entry of the gradient $\nabla_i f_\mu(x)$ must also be large. This still allows us to take a larger step in direction $\mathbf{e}_i$ than traditionally allowed by coordinate descent.

The above intuition is formalized in the next lemma, whose proof is by simple manipulation of Hessian. The first half of the lemma is the same as the traditional coordinate Lipschitz-smoothness property, but holds only conditionally; the second half is a salient characteristic of this work and requires the non-negativity of $A$. These smoothness properties will be crucial in applying gradient descent arguments in Section 3.3, and are the main motivation for us to adopt the $\|\cdot\|_A$ norm for our proposed algorithms.

**Lemma 2.7.** *Let $L \stackrel{\mathrm{def}}{=} \frac{4}{\mu}$. Then, for every $x \ge 0$, every $i \in [n]$, and every $\lambda \in \big[-\frac{1}{L\|A_{:i}\|_\infty}, \frac{1}{L\|A_{:i}\|_\infty}\big]$:*

*(a) If $|\nabla_i f_\mu(x)| \le 1$, then $\big|\nabla_i f_\mu(x+\lambda\mathbf{e}_i) - \nabla_i f_\mu(x)\big| \le L\|A_{:i}\|_\infty \cdot |\lambda|$ .*

*(b) If $\nabla_i f_\mu(x) \ge 1$, then $\nabla_i f_\mu(x+\lambda\mathbf{e}_i) \ge \Big(1 - \frac{\|A_{:i}\|_\infty L}{2}|\lambda|\Big)\nabla_i f_\mu(x)$ .*

*Proof of Lemma 2.7.* Using the fact that $\nabla_i f_\mu(x) > -1$ for all $x$, we have:

$$\begin{aligned}\Big|\log\frac{\nabla_i f_\mu(x+\lambda\mathbf{e}_i)+1}{\nabla_i f_\mu(x)+1}\Big| &\overset{①}{=} \Big|\int_0^\lambda \frac{\nabla^2_{ii} f_\mu(x+\nu\mathbf{e}_i)}{\nabla_i f_\mu(x+\nu\mathbf{e}_i)+1}d\nu\Big| \\ &\overset{②}{=} \frac{1}{\mu}\Big|\int_0^\lambda \frac{(A^T\mathrm{diag}\{p(x+\nu\mathbf{e}_i)\}A)_{ii}}{(A^T p(x+\nu\mathbf{e}_i))_i}d\nu\Big| \\ &\overset{③}{\le} \frac{\|A_{:i}\|_\infty}{\mu}|\lambda| \overset{④}{=} \frac{\|A_{:i}\|_\infty L}{4}|\lambda| \enspace .\end{aligned}$$

Above, ① holds because $\int_0^\lambda g'(\nu)d\nu = g(\lambda) - g(0)$ where $g(\nu) = \log(\nabla_i f_\mu(x+\nu\mathbf{e}_i)+1)$; ② holds according to Fact 2.6; ③ is because the numerator is $\sum_j A_{j,i}^2 p_j$ while the denominator is $\sum_j A_{j,i}p_j$; ④ holds because $L = \frac{4}{\mu}$. This immediately implies

$$e^{-\frac{\|A_{:i}\|_\infty L}{4}|\lambda|} \le \frac{\nabla_i f_\mu(x+\lambda\mathbf{e}_i)+1}{\nabla_i f_\mu(x)+1} \le e^{\frac{\|A_{:i}\|_\infty L}{4}|\lambda|}.$$

Our assumption on $\lambda$ implies $\frac{\|A_{:i}\|_\infty L}{4}|\lambda| \le \frac{1}{4}$, so that we can use the approximation $x \le e^x - 1 \le 1.2x$ over $x \in [-\frac{1}{4}, \frac{1}{4}]$. This yields the simpler bound:

$$-\frac{\|A_{:i}\|_\infty L}{4}|\lambda| \le \frac{\nabla_i f_\mu(x+\lambda\mathbf{e}_i) - \nabla_i f_\mu(x)}{\nabla_i f_\mu(x)+1} \le 1.2\frac{\|A_{:i}\|_\infty L}{4}|\lambda|.$$

(a) Assuming that $\nabla_i f_\mu(x) \in (-1, 1]$, we have:

$$\Big|\nabla_i f_\mu(x + \lambda \mathbf{e}_i) - \nabla_i f_\mu(x)\Big| \leq 2.4 \cdot \frac{\|A_{:i}\|_\infty L}{4}|\lambda| \leq \|A_{:i}\|_\infty L|\lambda| \enspace .$$

(b) Assuming $\nabla_i f_\mu(x) \geq 1$, we have

$$\nabla_i f_\mu(x + \lambda \mathbf{e}_i) \geq \nabla_i f_\mu(x) - \frac{\|A_{:i}\|_\infty L}{4}|\lambda|\Big(\nabla_i f_\mu(x) + 1\Big) \geq \Big(1 - \frac{\|A_{:i}\|_\infty L}{2}|\lambda|\Big)\nabla_i f_\mu(x) \enspace . \qquad \square$$

**Initialization.** Iterative methods require a starting point, and we use the following one

**Fact 2.8.** *Let* $x_i^{\mathsf{start}} \stackrel{\mathrm{def}}{=} \frac{1-\varepsilon/2}{n\|A_{:i}\|_\infty}$ *for each* $i \in [n]$*. Then,* $x^{\mathsf{start}} \in \Delta_{\mathsf{box}}$ *and* $f_\mu(x^{\mathsf{start}}) \leq -\frac{1-\varepsilon}{n}$*.*

*Proof.* Using the fact that $Ax^{\mathsf{start}} - \mathbb{1} \leq -\varepsilon/2 \cdot \mathbb{1}$, we compute $f_\mu(x^{\mathsf{start}})$ as follows:

$$\begin{aligned} f_\mu(x^{\mathsf{start}}) &= \mu \sum_j e^{\frac{1}{\mu}((Ax^{\mathsf{start}})_j - 1)} - \mathbb{1}^T x^{\mathsf{start}} \leq \mu \sum_j e^{\frac{-\varepsilon/2}{\mu}} - \frac{1-\varepsilon/2}{n} \\ &\leq \frac{\mu m}{(nm)^2} - \frac{1-\varepsilon/2}{n} \leq -\frac{1-\varepsilon}{n} \enspace . \end{aligned}$$

Above, we have used $\mathbb{1}^T x^{\mathsf{start}} \geq x_i^{\mathsf{start}} = \frac{1-\varepsilon/2}{n}$, where $i$ is the column s.t. $\|A_{:i}\|_\infty = 1$. $\square$

# 3 Our Packing LP Solver

Recall traditional (accelerated or not) gradient descent [29, 30] or coordinate descent [6, 18, 31] should not be applied directly to minimize $f_\mu$, because $f_\mu$ is not Lipschitz-smooth.

Our proposed algorithm `PacLPSolver` starts with some initial vector $\mathsf{x}_0 = \mathsf{y}_0 = x^{\mathsf{start}}$ (see Fact 2.8) and $\mathsf{z}_0 = 0$, and is divided into $T$ iterations. In each iteration $k$, it computes a weighted midpoint $\mathsf{x}_k \leftarrow \tau \mathsf{z}_{k-1} + (1-\tau)\mathsf{y}_{k-1}$ for some parameter $\tau \in (0, 1)$. This step is analogous to that in traditional accelerated coordinate descent [6, 18, 31]. We then compute $\mathsf{y}_k$ and $\mathsf{z}_k$ as follows.

We select $i \in [n]$ uniformly at random. Let $\xi_k^{(i)} = (0, \dots, 0, \mathbb{T}^{\mathsf{p}}(v), 0, \dots, 0)$ be the vector that is only non-zero at coordinate $i$, where $v = \nabla_i f_\mu(\mathsf{x}_k) \in [-1, \infty)$, and $\mathbb{T}^{\mathsf{p}}(v)$ is the thresholding function $\mathbb{T}^{\mathsf{p}}(v) \stackrel{\mathrm{def}}{=} \min\{v, 1\}$. We refer to $\xi_k^{(i)}$ as the *truncated gradient*.[10] Next,

- Perform a *mirror (descent) step* $\mathsf{z}_k \leftarrow \mathsf{z}_k^{(i)} \stackrel{\mathrm{def}}{=} \arg\min_{z \in \Delta_{\mathsf{box}}} \big\{\frac{1}{2}\|z - \mathsf{z}_{k-1}\|_A^2 + \langle n\alpha_k \xi_k^{(i)}, z\rangle\big\}$ for some parameter $\alpha_k \ll 1/n$ to be chosen later.
- Perform a *gradient (descent) step* $\mathsf{y}_k \leftarrow \mathsf{y}_k^{(i)} \stackrel{\mathrm{def}}{=} \mathsf{x}_k + \frac{1}{n\alpha_k L}(\mathsf{z}_k^{(i)} - \mathsf{z}_{k-1})$.

This finishes the description of our `PacLPSolver`.

**Remark 3.1.** We use the superscript $^{(i)}$ on $\xi_k^{(i)}$, $\mathsf{y}_k^{(i)}$ and $\mathsf{z}_k^{(i)}$ to emphasize that the value depends on the choice of $i$. We use generic parameters $\tau, \alpha_k, T$ in the above description and their precise values are presented in Algorithm 1.

Our update on $\mathsf{y}_k$ is a "gradient descent step" because we shall prove that it strictly decreases the objective (i.e., $f_\mu(\mathsf{x}_k) - f_\mu(\mathsf{y}_k^{(i)}) \geq 0$). Our update on $\mathsf{z}_k$ is a "mirror descent step" because we shall apply standard mirror descent analysis [13] to it. We explicitly describe how to implement the mirror step (its proof is straightforward by computing the gradient):

[10] A similar gradient truncation was developed in our prior work [2], but for a different purpose (to ensure parallelism) and not applied to coordinate gradient. The truncation idea of this paper also inspired later works in matrix scaling [7] and in SDP [5].

**Algorithm 1** $\texttt{PacLPSolver}(A, x^{\mathsf{start}}, \varepsilon)$

**Input:** $A \in \mathbb{R}_{\geq 0}^{m \times n}, x^{\mathsf{start}} \in \Delta_{\mathsf{box}}, \varepsilon \in (0, 1/30]$.

**Output:** $x \in \Delta_{\mathsf{box}}$. ▷ recall $\Delta_{\mathsf{box}} \stackrel{\text{def}}{=} \{x \in \mathbb{R}^n : x_i \in \big[0, \frac{1}{\|A_{:i}\|_\infty}\big]\}$

$\mu \leftarrow \frac{\varepsilon}{4\log(nm/\varepsilon)}$, $L \leftarrow \frac{4}{\mu}$, $\tau \leftarrow \frac{1}{3 \cdot nL}$ and $\alpha_0 \leftarrow \frac{1}{nL}$. ▷ parameters

$T \leftarrow \lceil 3nL \log(1/\varepsilon) \rceil = O(n \cdot \frac{\log(nm/\varepsilon) \cdot \log(1/\varepsilon)}{\varepsilon})$. ▷ number of iterations

$\mathsf{x}_0 = \mathsf{y}_0 \leftarrow x^{\mathsf{start}}$, $\mathsf{z}_0 \leftarrow 0$.

**for** $k \leftarrow 1$ **to** $T$ **do**

  $\alpha_k \leftarrow \frac{1}{1-\tau} \alpha_{k-1}$

  $\mathsf{x}_k \leftarrow \tau \mathsf{z}_{k-1} + (1-\tau) \mathsf{y}_{k-1}$.

  Randomly select $i \in [n]$ uniformly at random.

  Define vector $\xi_k^{(i)}$ to be all-zero except at coordinate $i$, where $\xi_{k,i}^{(i)} = \min\{1, \nabla_i f_\mu(\mathsf{x}_k)\}$.

  $\mathsf{z}_k \leftarrow \mathsf{z}_k^{(i)} \stackrel{\text{def}}{=} \arg\min_{z \in \Delta_{\mathsf{box}}} \big\{ \frac{1}{2} \|z - \mathsf{z}_{k-1}\|_A^2 + \langle n\alpha_k \xi_k^{(i)}, z \rangle \big\}$. ▷ See Proposition 3.2

  $\mathsf{y}_k \leftarrow \mathsf{y}_k^{(i)} \stackrel{\text{def}}{=} \mathsf{x}_k + \frac{1}{n\alpha_k L}(\mathsf{z}_k^{(i)} - \mathsf{z}_{k-1})$.

**end for**

**return** $\mathsf{y}_T$.

**Proposition 3.2.** *If $\Delta_{\mathsf{box}} = \{x \in \mathbb{R}^n : x_i \in \big[0, \frac{C}{\|A_{:i}\|_\infty}\big]\}$ for some constant $C > 0$, the minimizer $z = \arg\min_{z \in \Delta_{\mathsf{box}}} \big\{ \frac{1}{2}\|z - \mathsf{z}_{k-1}\|_A^2 + \langle \delta \mathbf{e}_i, z \rangle \big\}$ for any $\delta \in \mathbb{R}$ and basis vector $\mathbf{e}_i$ can be computed as follows:*

1. *$z \leftarrow \mathsf{z}_{k-1}$.*
2. *$z_i \leftarrow z_i - \delta / \|A_{:i}\|_\infty$.*
3. *If $z_i < 0$, then $z_i \leftarrow 0$; if $z_i > C/\|A_{:i}\|_\infty$, $z_i \leftarrow C/\|A_{:i}\|_\infty$.*
4. *Return $z$.*

We also point out that

**Lemma 3.3.** *Each iteration of* `PacLPSolver` *can be implemented to run in expected $O(N/n)$ time. The total expected running time is $O(TN/n)$.*

Lemma 3.3 is not hard to prove, but anyways included in Appendix E. It follows from standard implementation tricks which compute $\mathsf{x}_k$ and $\mathsf{y}_k$ only *implicitly*: that is to express $\mathsf{x}_k$ and $\mathsf{y}_k$ as linear combinations of two less-frequently-updated vectors.

## 3.1 Convergence Statement

In this section, we focus on proving the following main theorem.

**Theorem 3.4.** $\texttt{PacLPSolver}(A, x^{\mathsf{start}}, \varepsilon)$ *outputs some $\mathsf{y}_T$ satisfying*

$$\mathbf{E}[f_\mu(\mathsf{y}_T)] \leq -(1 - 3\varepsilon)\mathsf{OPT} \ .$$

It is straightforward to use Markov's bound to turn Theorem 3.4 into a probabilistic one

**Corollary 3.5.** *With probability at least $2/3$, the output $\mathsf{y}_T = \texttt{PacLPSolver}(A, x^{\mathsf{start}}, \varepsilon)$ satisfies that $\frac{\mathsf{y}_T}{1+\varepsilon}$ is a $(1 - O(\varepsilon))$ approximate solution to the packing LP program. The expected running time is $O(\frac{\log(nm/\varepsilon)\log(1/\varepsilon)}{\varepsilon} N)$.*

*Proof of Corollary 3.5.* Since for every $x \in \Delta_{\mathsf{box}}$ it satisfies $f_\mu(x) \geq -(1+\varepsilon)\mathsf{OPT}$ according to Proposition 2.4.b, we obtain that $f_\mu(y_T) + (1+\varepsilon)\mathsf{OPT}$ is a random variable that is non-negative, whose expectation $\mathbf{E}[f_\mu(y_T) + (1+\varepsilon)\mathsf{OPT}] \le 4\varepsilon$ according to Theorem 3.4. By Markov bound, with at least probability 2/3, we obtain some $y_T$ satisfying $f_\mu(y_T) \le -(1-11\varepsilon)\mathsf{OPT}$, which yields a $(1-O(\varepsilon))$ approximate solution to packing LP according to Proposition 2.4.c. The running time follows from Lemma 3.3. □

Before we prove prove Theorem 3.4 in subsequent subsections, let us first point out that our iterates $\mathsf{x}_k, \mathsf{y}_k, z_k$ never leave the bounding box $\Delta_{\mathsf{box}}$:

**Lemma 3.6.** *We have* $\mathsf{x}_k, \mathsf{y}_k, \mathsf{z}_k \in \Delta_{\mathsf{box}}$ *for all* $k = 0, 1, \dots, T$.

(The proof of Lemma 3.6 is included in Appendix A, and the main technique already appeared in randomized coordinate descent [18].)

## 3.2 Step 1: Mirror Descent Guarantee

Following almost classical analysis of mirror descent (cf. textbook [13]), our update $\mathsf{z}_k^{(i)} = \arg\min_{z\in\Delta_{\mathsf{box}}} \big\{\frac{1}{2}\|z-\mathsf{z}_{k-1}\|_A^2 + \langle n\alpha_k \xi_k^{(i)}, z\rangle\big\}$ satisfies

**Lemma 3.7** (mirror descent)**.** *For every* $u \in \Delta_{\mathsf{box}}$, *it satisfies*

$$\big\langle n\alpha_k \xi_k^{(i)}, \mathsf{z}_{k-1} - u\big\rangle \le n^2\alpha_k^2 L \cdot \big\langle \xi_k^{(i)}, \mathsf{x}_k - \mathsf{y}_k^{(i)}\big\rangle + \frac{1}{2}\|\mathsf{z}_{k-1} - u\|_A^2 - \frac{1}{2}\|\mathsf{z}_k^{(i)} - u\|_A^2 \ .$$

*Proof.* Denoting by $V_a(b) = \frac{1}{2}\|b-a\|_A^2$ as a function of $b \in \Delta_{\mathsf{box}}$ parameterized at $a \in \Delta_{\mathsf{box}}$, we have $\nabla_i V_a(b) = \|A_{:i}\|_\infty \cdot (a_i - b_i)$. In the optimization language, $V_a(b)$ is the Bregman divergence of the $\|\cdot\|_A^2$ regularizer [13]. We derive the following sequence of inequalities:

$$\begin{aligned}
\big\langle n\alpha_k \xi_k^{(i)}, \mathsf{z}_{k-1} - u\big\rangle &= \big\langle n\alpha_k \xi_k^{(i)}, \mathsf{z}_{k-1} - \mathsf{z}_k^{(i)}\big\rangle + \big\langle n\alpha_k \xi_k^{(i)}, \mathsf{z}_k^{(i)} - u\big\rangle \\
&\overset{①}{\le} \big\langle n\alpha_k \xi_k^{(i)}, \mathsf{z}_{k-1} - \mathsf{z}_k^{(i)}\big\rangle + \big\langle -\nabla V_{\mathsf{z}_{k-1}}(\mathsf{z}_k^{(i)}), \mathsf{z}_k^{(i)} - u\big\rangle \\
&\overset{②}{=} \big\langle n\alpha_k \xi_k^{(i)}, \mathsf{z}_{k-1} - \mathsf{z}_k^{(i)}\big\rangle - \frac{1}{2}\|\mathsf{z}_{k-1} - \mathsf{z}_k^{(i)}\|_A^2 + \frac{1}{2}\|\mathsf{z}_{k-1} - u\|_A^2 - \frac{1}{2}\|\mathsf{z}_k^{(i)} - u\|_A^2 \\
&\overset{③}{=} n^2\alpha_k^2 L\Big(\big\langle \xi_k^{(i)}, \mathsf{x}_k - \mathsf{y}_k\big\rangle - \frac{L}{2}\|\mathsf{x}_k - \mathsf{y}_k\|_A^2\Big) + \frac{1}{2}\|\mathsf{z}_{k-1} - u\|_A^2 - \frac{1}{2}\|\mathsf{z}_k^{(i)} - u\|_A^2 \\
&\le n^2\alpha_k^2 L \cdot \big\langle \xi_k^{(i)}, \mathsf{x}_k - \mathsf{y}_k\big\rangle + \frac{1}{2}\|\mathsf{z}_{k-1} - u\|_A^2 - \frac{1}{2}\|\mathsf{z}_k^{(i)} - u\|_A^2 \ .
\end{aligned}$$

Above, ① is due to the minimality of $\mathsf{z}_k^{(i)} = \arg\min_{z\in\Delta_{\mathsf{box}}} \big\{V_{\mathsf{z}_{k-1}}(z) + \big\langle n\alpha_k \xi_k^{(i)}, z\big\rangle\big\}$, which implies that $\big\langle \nabla V_{\mathsf{z}_{k-1}}(\mathsf{z}_k^{(i)}) + n\alpha\xi_k^{(i)}, u - \mathsf{z}_k^{(i)}\big\rangle \ge 0$ for all $u \in \Delta_{\mathsf{box}}$. Equality ② can be checked for every coordinate $\ell \in [n]$ as follows:

$$\begin{aligned}
-\nabla_\ell V_{\mathsf{z}_{k-1}}(\mathsf{z}_k^{(i)}) \cdot (\mathsf{z}_{k,\ell}^{(i)} - u_\ell) &= \|A_{:i}\|_\infty (\mathsf{z}_{k-1,\ell} - \mathsf{z}_{k,\ell}^{(i)}) \cdot (\mathsf{z}_{k,\ell}^{(i)} - u_\ell) \\
&= \|A_{:i}\|_\infty \Big( -\frac{1}{2}(\mathsf{z}_{k-1,\ell} - \mathsf{z}_{k,\ell}^{(i)})^2 + \frac{1}{2}(u_\ell - \mathsf{z}_{k-1,\ell})^2 - \frac{1}{2}(\mathsf{z}_{k,\ell}^{(i)} - u_\ell)^2\Big) \ .
\end{aligned}$$

③ is by our choice of $\mathsf{y}_k$ which satisfies that $\mathsf{z}_{k-1} - \mathsf{z}_k^{(i)} = n\alpha_k L(\mathsf{x}_k - \mathsf{y}_k^{(i)})$. □

In addition, as a simple corollary of Proposition 3.2, we have the following fact

**Fact 3.8.** $|\mathsf{z}_{k,i}^{(i)} - \mathsf{z}_{k-1,i}| \le \frac{n\alpha_k |\xi_{k,i}^{(i)}|}{\|A_{:i}\|_\infty}$ *and* $|\mathsf{y}_{k,i}^{(i)} - \mathsf{x}_{k,i}| = \frac{1}{n\alpha_k L}|\mathsf{z}_{k,i}^{(i)} - \mathsf{z}_{k-1,i}| \le \frac{|\xi_{k,i}^{(k)}|}{L\|A_{:i}\|_\infty} \le \frac{1}{L\|A_{:i}\|_\infty}$. *If* $\xi_{k,i}^{(i)} \ge 0$, *then* $\mathsf{z}_{k,i}^{(i)} \le \mathsf{z}_{k-1,i}$ *and* $\mathsf{y}_{k,i}^{(i)} \le \mathsf{x}_{k,i}$; *if* $\xi_{k,i}^{(i)} \le 0$, *then* $\mathsf{z}_{k,i}^{(i)} \ge \mathsf{z}_{k-1,i}$ *and* $\mathsf{y}_{k,i}^{(i)} \ge \mathsf{x}_{k,i}$.

### 3.3 Step 2: Gradient Descent Guarantee

We call our update $\mathsf{y}_k^{(i)} \leftarrow \mathsf{x}_k + \frac{1}{n\alpha_k L}(\mathsf{z}_k^{(i)} - \mathsf{z}_{k-1})$ a gradient descent step, because the following lemma guarantees $f_\mu(\mathsf{y}_k^{(i)}) \leq f_\mu(\mathsf{x}_k)$, that is, the objective only decreases; moreover, the objective decreases at least by $\frac{1}{2}\langle \nabla f_\mu(\mathsf{x}_k), \mathsf{x}_k - \mathsf{y}_k^{(i)} \rangle$.

**Lemma 3.9** (gradient descent)**.** *We have* $f_\mu(\mathsf{x}_k) - f_\mu(\mathsf{y}_k^{(i)}) \geq \frac{1}{2}\langle \nabla f_\mu(\mathsf{x}_k), \mathsf{x}_k - \mathsf{y}_k^{(i)} \rangle \geq 0$.

This Lemma 3.9, which is characteristic of the PC-LP setting, is strong in the following sense. Even though the update $y_k^{(i)}$ only depends on the truncated gradient $\xi_k^{(i)}$, the progress we make is a function of the true gradient $\nabla_i f_\mu(\mathsf{x}_k)$, including the large component that was discarded. This is possible because the smoothness guarantee of Lemma 2.7.b allows us to take a long coordinate step even though $f_\mu(x)$ is not Lipschitz-smooth.

*Proof of Lemma 3.9.* Using Fact 3.8, we write $\mathsf{y}_k^{(i)} = \mathsf{x}_k + s\lambda \mathbf{e}_i$ for some $s \in \{-1, +1\}$ and step length $\lambda \in \big[0, \frac{1}{L\|A_{:i}\|_\infty}\big]$. We first focus on the case $\nabla_i f_\mu(\mathsf{x}_k) \in [-1, 1]$ so $\xi_{k,i}^{(i)} = \nabla_i f_\mu(\mathsf{x}_k)$.

$$f_\mu(\mathsf{x}_k) - f_\mu(\mathsf{y}_k^{(i)}) = f_\mu(\mathsf{x}_k) - f_\mu(\mathsf{x}_k + s\lambda \mathbf{e}_i) = -s\int_0^\lambda \Big(\nabla_i f_\mu(\mathsf{x}_k + s\chi \mathbf{e}_i)\Big) d\chi$$
$$\overset{①}{\geq} s\int_0^\lambda \Big(-\nabla_i f_\mu(\mathsf{x}_k) - L\|A_{:i}\|_\infty \cdot s\chi\Big) d\chi = -\nabla_i f_\mu(\mathsf{x}_k)\cdot s\lambda - \frac{L\|A_{:i}\|_\infty}{2}\cdot \lambda^2$$
$$\overset{②}{\geq} -\nabla_i f_\mu(\mathsf{x}_k)\cdot s\lambda - \frac{L\|A_{:i}\|_\infty}{2}\cdot\lambda\cdot\frac{|\xi_{k,i}^{(k)}|}{L\|A_{:i}\|_\infty} \overset{③}{=} -\frac{1}{2}\langle \nabla f_\mu(\mathsf{x}_k), \mathsf{y}_k^{(i)} - \mathsf{x}_k\rangle \ .$$

Above, ① uses Lemma 2.7.a, ② uses Fact 3.8, and ③ uses $|\xi_{k,i}^{(k)}| = -s\nabla_i f_\mu(\mathsf{x}_k)$ (see also Fact 3.8). Next, we turn to the case of $\nabla_i f_\mu(\mathsf{x}_k) > 1$. In this case, we have $s = -1$ and

$$f_\mu(\mathsf{x}_k) - f_\mu(\mathsf{y}_k^{(i)}) = f_\mu(\mathsf{x}_k) - f_\mu(\mathsf{x}_k - \lambda \mathbf{e}_i) = \int_0^\lambda \nabla_i f_\mu(\mathsf{x}_k - \chi \mathbf{e}_i) d\chi$$
$$\overset{①}{\geq} \int_0^\lambda \Big(1 - \frac{\|A_{:i}\|_\infty L}{2}\chi\Big)\nabla_i f_\mu(x) d\chi \overset{②}{\geq} \int_0^\lambda \frac{1}{2}\nabla_i f_\mu(x) d\chi = \frac{1}{2}\langle \nabla f_\mu(\mathsf{x}_k), \mathsf{x}_k - \mathsf{y}_k^{(i)}\rangle \ .$$

Above, ① uses Lemma 2.7.b and ② uses $\chi \leq \lambda \leq \frac{1}{L\|A_{:i}\|_\infty}$. Finally, we have $\langle \nabla f_\mu(\mathsf{x}_k), \mathsf{x}_k - \mathsf{y}_k^{(i)}\rangle \geq 0$ because $\nabla_i f_\mu(\mathsf{x}_k)$ and $\mathsf{x}_{k,i} - \mathsf{y}_{k,i}^{(i)}$ have the same sign, and $\mathsf{x}_{k,\ell} = \mathsf{y}_{k,\ell}^{(i)}$ for $\ell \neq i$. □

### 3.4 Step 3: Putting All Together

We denote by $\eta_k^{(i)} \in \mathbb{R}_{\geq 0}^n$ the vector that is only non-zero at coordinate $i$, and satisfies $\eta_{k,i}^{(i)} = \nabla_i f_\mu(\mathsf{x}_k) - \xi_{k,i}^{(i)} \in [0, \infty)$. In other words, the full gradient

$$\nabla f_\mu(\mathsf{x}_k) = \mathbf{E}_i[(0, \dots, n\nabla_i f_\mu(\mathsf{x}_k), \dots, 0)] = \mathbf{E}_i[n\eta_k^{(i)} + n\xi_k^{(i)}]$$

can be (in expectation) decomposed into the a large non-negative component $\eta_k^{(i)} \in [0, \infty)^n$ and a truncated component $\xi_k^{(i)} \in [-1, 1]^n$. Recall that $\eta_k^{(i)}$ did not contribute to the descent steps (see Line 9 of `PacLPSolver`). Now, for any $u \in \Delta_{\mathsf{box}}$, we can use a basic convexity argument and the mirror descent lemma to compute that

$$\alpha_k(f_\mu(\mathsf{x}_k) - f_\mu(u)) \leq \big\langle \alpha_k \nabla f_\mu(\mathsf{x}_k), \mathsf{x}_k - u\big\rangle$$
$$= \big\langle \alpha_k \nabla f_\mu(\mathsf{x}_k), \mathsf{x}_k - \mathsf{z}_{k-1}\big\rangle + \big\langle \alpha_k \nabla f_\mu(\mathsf{x}_k), \mathsf{z}_{k-1} - u\big\rangle$$
$$= \big\langle \alpha_k \nabla f_\mu(\mathsf{x}_k), \mathsf{x}_k - \mathsf{z}_{k-1}\big\rangle + \mathbf{E}_i\Big[\big\langle n\alpha_k \eta_k^{(i)}, \mathsf{z}_{k-1} - u\big\rangle + \big\langle n\alpha_k \xi_k^{(i)}, \mathsf{z}_{k-1} - u\big\rangle\Big]$$

$$\overset{①}{=} \frac{(1-\tau)\alpha_k}{\tau}\langle \nabla f_\mu(\mathsf{x}_k), \mathsf{y}_{k-1} - \mathsf{x}_k\rangle + \mathbf{E}_i\left[\langle n\alpha_k \eta_k^{(i)}, \mathsf{z}_{k-1} - u\rangle + \langle n\alpha_k \xi_k^{(i)}, \mathsf{z}_{k-1} - u\rangle\right] \tag{3.1}$$

$$\overset{②}{\le} \frac{(1-\tau)\alpha_k}{\tau}(f_\mu(\mathsf{y}_{k-1}) - f_\mu(\mathsf{x}_k))$$

$$+ \mathbf{E}_i\Big[\boxed{\langle n\alpha_k \eta_k^{(i)}, \mathsf{z}_{k-1} - u\rangle + n^2\alpha_k^2 L \cdot \langle \xi_k^{(i)}, \mathsf{x}_k - \mathsf{y}_k^{(i)}\rangle} + \frac{1}{2}\|\mathsf{z}_{k-1} - u\|_A^2 - \frac{1}{2}\|\mathsf{z}_k^{(i)} - u\|_A^2\Big] \tag{3.2}$$

Above, ① is because $\mathsf{x}_k = \tau\mathsf{z}_{k-1} + (1-\tau)\mathsf{y}_{k-1}$, which implies that $\tau(\mathsf{x}_k - \mathsf{z}_{k-1}) = (1-\tau)(\mathsf{y}_{k-1} - \mathsf{x}_k)$. ② uses convexity and Lemma 3.7. This above computation is motivated by [3], and as we shall see below, it allows one to linearly couple gradient and mirror steps.

Intuitively, the first term in the box of (3.2) is the loss introduced by the large gradient $\eta_k^{(i)}$. This part was truncated so did not contribute to the mirror step. The second term in the box is the loss introduced by mirror descent on the small gradient $\xi_k^{(i)}$ in Lemma 3.7.

Now comes an important observation. As shown by Lemma 3.10 below, the performance of the gradient step —that is, the objective decrease of $f_\mu(\mathsf{x}_k) - f_\mu(\mathsf{y}_k^{(i)})$— is at least proportional to the total loss incurred in the box. Intuitively, this means that the progress in the gradient step is so large that it outweighs not only the loss from mirror descent (as is typical in accelerated gradient analyses [3, 30]) but also the loss term introduced by $\eta_k^{(i)}$.

**Lemma 3.10** (gradient descent total guarantee)**.** *For every $u \ge 0$,*

$$\langle n\alpha_k \eta_k^{(i)}, \mathsf{z}_{k-1} - u\rangle + n^2\alpha_k^2 L \cdot \langle \xi_k^{(i)}, \mathsf{x}_k - \mathsf{y}_k^{(i)}\rangle \le 3n\alpha_k L \cdot (f_\mu(\mathsf{x}_k) - f_\mu(\mathsf{y}_k^{(i)})) \ .$$

The proof of Lemma 3.10 is a careful case analysis and several simple applications of Lemma 3.9. We remark that to properly upper bound $\langle n\alpha_k \eta_k^{(i)}, \mathsf{z}_{k-1} - u\rangle$, one needs to have some good upper bound the coordinates of $\mathsf{z}_{k-1}$. This is exactly the place we need our redundant constraint which ensures $\mathsf{z}_{k-1,i} \le \frac{1}{\|A_{:i}\|_\infty}$ (see Remark 2.5).

*Proof of Lemma 3.10.* There are three possibilities:

- If $\eta_{k,i}^{(i)} = 0$, then we must have $\xi_{k,i}^{(i)} = \nabla_i f_\mu(\mathsf{x}_k) \in [-1, 1]$. Lemma 3.9 implies

$$\langle n\alpha_k \eta_k^{(i)}, \mathsf{z}_{k-1} - u\rangle + n^2\alpha_k^2 L \cdot \langle \xi_k^{(i)}, \mathsf{x}_k - \mathsf{y}_k^{(i)}\rangle$$
$$= n^2\alpha_k^2 L \cdot \langle \nabla f_\mu(\mathsf{x}_k), \mathsf{x}_k - \mathsf{y}_k^{(i)}\rangle \le 2n^2\alpha_k^2 L \cdot (f_\mu(\mathsf{x}_k) - f_\mu(\mathsf{y}_k^{(i)}))$$

- If $\eta_{k,i}^{(i)} > 0$ and $\mathsf{z}_{k,i}^{(i)} > 0$, then we precisely have $\mathsf{z}_{k,i}^{(i)} = \mathsf{z}_{k-1,i} - \frac{n\alpha_k}{\|A_{:i}\|_\infty}$ (see Proposition 3.2), and accordingly $\mathsf{y}_{k,i}^{(i)} = \mathsf{x}_{k,i} - \frac{1}{L\|A_{:i}\|_\infty} < \mathsf{x}_{k,i}$. In this case,

$$\langle n\alpha_k \eta_k^{(i)}, \mathsf{z}_{k-1} - u\rangle + n^2\alpha_k^2 L \cdot \langle \xi_k^{(i)}, \mathsf{x}_k - \mathsf{y}_k^{(i)}\rangle$$
$$\overset{①}{\le} n\alpha_k \cdot \nabla_i f_\mu(\mathsf{x}_k) \cdot \frac{1}{\|A_{:i}\|_\infty} + n^2\alpha_k^2 L \cdot \langle \xi_k^{(i)}, \mathsf{x}_k - \mathsf{y}_k^{(i)}\rangle$$
$$\overset{②}{<} n\alpha_k \cdot \nabla_i f_\mu(\mathsf{x}_k) \cdot \frac{1}{\|A_{:i}\|_\infty} + n^2\alpha_k^2 L \cdot \langle \nabla f_\mu(\mathsf{x}_k), \mathsf{x}_k - \mathsf{y}_k^{(i)}\rangle$$
$$\overset{③}{=} n\alpha_k L \cdot \langle \nabla f_\mu(\mathsf{x}_k), \mathsf{x}_k - \mathsf{y}_k^{(i)}\rangle + n^2\alpha_k^2 L \cdot \langle \nabla f_\mu(\mathsf{x}_k), \mathsf{x}_k - \mathsf{y}_k^{(i)}\rangle$$
$$\overset{④}{\le} \left(2n\alpha_k L + 2n^2\alpha_k^2 L\right) \cdot (f_\mu(\mathsf{x}_k) - f_\mu(\mathsf{y}_k^{(i)})) \ .$$

  Above, ① follows from the fact that $\mathsf{z}_{k-1} \in \Delta_{\mathsf{box}}$ and therefore $\mathsf{z}_{k-1,i} \le \frac{1}{\|A_{:i}\|_\infty}$ by the definition of $\Delta_{\mathsf{box}}$, and $u \ge 0$; ② follows from the fact that $\mathsf{x}_k$ and $\mathsf{y}_k^{(i)}$ are only different at coordinate $i$, and $\xi_{k,i}^{(i)} = 1 < \nabla_i f_\mu(\mathsf{x}_k)$ (since $\eta_{k,i}^{(i)} > 0$); ③ follows from the fact that $\mathsf{y}_k^{(i)} = \mathsf{x}_k - \frac{\mathbf{e}_i}{L\|A_{:i}\|_\infty}$; and ④ uses Lemma 3.9.

- If $\eta_{k,i}^{(i)} > 0$ and $\mathsf{z}_{k,i}^{(i)} = 0$, then we have

  $$\begin{aligned}&\big\langle n\alpha_k \eta_k^{(i)}, \mathsf{z}_{k-1} - u\big\rangle + n^2\alpha_k^2 L \cdot \big\langle \xi_k^{(i)}, \mathsf{x}_k - \mathsf{y}_k^{(i)}\big\rangle \\ &\overset{①}{\leq} \big(n\alpha_k \nabla_i f_\mu(\mathsf{x}_k) \cdot \mathsf{z}_{k-1,i}\big) + n^2\alpha_k^2 L \cdot \big\langle \nabla f_\mu(\mathsf{x}_k), \mathsf{x}_k - \mathsf{y}_k^{(i)}\big\rangle \\ &\overset{②}{=} \big\langle n\alpha_k \nabla f_\mu(\mathsf{x}_k), \mathsf{z}_{k-1} - \mathsf{z}_k^{(i)}\big\rangle + n^2\alpha_k^2 L \cdot \big\langle \nabla f_\mu(\mathsf{x}_k), \mathsf{x}_k - \mathsf{y}_k^{(i)}\big\rangle \\ &\overset{③}{=} n^2\alpha_k^2 L \cdot \big\langle \nabla f_\mu(\mathsf{x}_k), \mathsf{x}_k - \mathsf{y}_k^{(i)}\big\rangle + n^2\alpha_k^2 L \cdot \big\langle \nabla f_\mu(\mathsf{x}_k), \mathsf{x}_k - \mathsf{y}_k^{(i)}\big\rangle \overset{④}{\leq} 4n^2\alpha_k^2 L \cdot \big(f_\mu(\mathsf{x}_k) - f_\mu(\mathsf{y}_k^{(i)})\big) \ .\end{aligned}$$

  Above, ① is because $u \geq 0$, $\nabla_i f_\mu(\mathsf{x}_k) = \eta_{k,i}^{(i)} + 1 > \eta_{k,i}^{(i)}$ and $\nabla_i f_\mu(\mathsf{x}_k) > \xi_{k,i}^{(i)}$; ② uses the assumption that $\mathsf{z}_{k,i}^{(i)} = 0$ and the fact that $\mathsf{z}_{k-1,\ell} = \mathsf{z}_{k,\ell}^{(i)}$ for every $\ell \neq i$; ③ is from our choice of $\mathsf{y}_k$ which satisfies that $\mathsf{z}_{k-1} - \mathsf{z}_k^{(i)} = n\alpha_k L(\mathsf{x}_k - \mathsf{y}_k^{(i)})$; and ④ uses Lemma 3.9.

Combining the three cases, and using the fact that $f_\mu(\mathsf{x}_k) - f_\mu(\mathsf{y}_k^{(i)}) \geq 0$, we conclude that

$$\begin{aligned}\big\langle n\alpha_k \eta_k^{(i)}, \mathsf{z}_{k-1} - u\big\rangle + n^2\alpha_k^2 L \cdot \big\langle \xi_k^{(i)}, \mathsf{x}_k - \mathsf{y}_k^{(i)}\big\rangle &\leq (2n\alpha_k L + 4n^2\alpha_k^2 L) \cdot \big(f_\mu(\mathsf{x}_k) - f_\mu(\mathsf{y}_k^{(i)})\big) \\ &\leq 3n\alpha_k L \cdot \big(f_\mu(\mathsf{x}_k) - f_\mu(\mathsf{y}_k^{(i)})\big) \ .\end{aligned}$$

Above, the last inequality uses our choice of $\alpha_k$, which implies $n\alpha_k \leq n\alpha_T = \frac{1}{\varepsilon L} \leq \frac{1}{4}$. □

Plugging Lemma 3.10 back to (3.2), we have

$$\begin{aligned}&\alpha_k(f_\mu(\mathsf{x}_k) - f_\mu(u)) \leq \big\langle \alpha_k \nabla f_\mu(\mathsf{x}_k), \mathsf{x}_k - u\big\rangle \\ &\overset{①}{\leq} \frac{(1-\tau)\alpha_k}{\tau}(f_\mu(\mathsf{y}_{k-1}) - f_\mu(\mathsf{x}_k)) + \mathbf{E}_i\Big[3n\alpha_k L \cdot \big(f_\mu(\mathsf{x}_k) - f_\mu(\mathsf{y}_k^{(i)})\big) + \frac{1}{2}\|\mathsf{z}_{k-1} - u\|_A^2 - \frac{1}{2}\|\mathsf{z}_k - u\|_A^2\Big] \\ &\overset{②}{\leq} \alpha_k f_\mu(\mathsf{x}_k) + \big(3n\alpha_k L - \alpha_k\big) f_\mu(\mathsf{y}_{k-1}) + \mathbf{E}_i\Big[-3n\alpha_k L \cdot f_\mu(\mathsf{y}_k^{(i)}) + \frac{1}{2}\|\mathsf{z}_{k-1} - u\|_A^2 - \frac{1}{2}\|\mathsf{z}_k - u\|_A^2\Big] \ .\end{aligned} \tag{3.3}$$

Above, ① uses Lemma 3.10; and ② is because we have chosen $\tau$ to satisfy $\frac{1}{\tau} = 3nL$.

Next, recall that we have picked $\alpha_k$ so that $(3nL-1)\alpha_k = 3nL\cdot\alpha_{k-1}$ in Algorithm 1. Telescoping (3.3) for $k = 1, \dots, T$ and choosing $u^* = (1-\varepsilon/2)x^*$, we have

$$-\sum_{k=1}^{T} \alpha_k f_\mu(u^*) \leq 3 f_\mu(\mathsf{y}_0) - 3n\alpha_T L \cdot \mathbf{E}[f_\mu(\mathsf{y}_T)] + \|\mathsf{z}_0 - u^*\|_A^2 \leq -3n\alpha_T L \cdot \mathbf{E}[f_\mu(\mathsf{y}_T)] + \mathsf{OPT} \ .$$

Here, the second inequality is due to $f_\mu(\mathsf{y}_0) = f_\mu(x^{\mathsf{start}}) \leq 0$ from Fact 2.8, and the fact that

$$\|\mathsf{z}_0 - u^*\|_A^2 = \|u^*\|_A^2 = \sum_{i=1}^{n} (u_i^*)^2 \cdot \|A_{:i}\|_\infty \leq \sum_{i=1}^{n} (x_i^*)^2 \cdot \|A_{:i}\|_\infty \leq \sum_{i=1}^{n} x_i^* = \mathsf{OPT} \ .$$

Finally, using the fact that $\sum_{k=1}^{T} \alpha_k = \alpha_T \cdot \sum_{k=0}^{T-1} \big(1 - \frac{1}{3nL}\big)^k = 3n\alpha_T L\big(1 - (1 - \frac{1}{3nL})^T\big)$, we rearrange and obtain that

$$\mathbf{E}[f_\mu(\mathsf{y}_T)] \leq \frac{\sum_k \alpha_k}{3n\alpha_T L} f_\mu(u^*) + \frac{1}{3n\alpha_T L}\mathsf{OPT} = \big(1 - (1 - \frac{1}{3nL})^T\big) f_\mu(u^*) + \frac{1}{3n\alpha_T L}\mathsf{OPT} \ .$$

We choose $T = \lceil 3nL \log(1/\varepsilon) \rceil$ so that $\frac{1}{n\alpha_T L} = (1 - \frac{1}{3nL})^T \leq \varepsilon$. Combining this with the fact that $f_\mu(u^*) \leq -(1-\varepsilon)\mathsf{OPT} < 0$ (see Proposition 2.4.a), we obtain

$$\mathbf{E}[f_\mu(\mathsf{y}_T)] \leq (1-\varepsilon) f_\mu(u^*) + \varepsilon/3 \cdot \mathsf{OPT} < -(1-3\varepsilon)\mathsf{OPT} \ .$$

Therefore, we have finished proving Theorem 3.4. □

# 4 Relaxation of the Covering Linear Program

Since PacLPSolver gives only an approximate packing LP solution, we cannot infer from it a dual covering LP solution. Therefore, we have to work on a relaxed version of covering LP directly. For input matrix $A \in \mathbb{R}_{\geq 0}^{m \times n}$, we rewrite the covering LP problem (1.2) as follows in order to be notationally close to packing LP:

$$\min_{x \geq 0} \{ \mathbb{1}^T x \,:\, Ax \geq \mathbb{1} \} \ . \tag{4.1}$$

We denote by $\mathsf{OPT}$ the optimal value to this LP, and by $x^*$ any of its optimal solutions. We say that $x$ is a $(1+\varepsilon)$-approximation for the covering LP if $Ax \geq \mathbb{1}$ and $\mathbb{1}^T x \leq (1+\varepsilon)\mathsf{OPT}$.

Again, we use indices $i \in [n]$ for the columns of $A$, and indices $j \in [m]$ for the rows of $A$. We denote by $A_{:i}$ the $i$-th column vector of $A$, and $A_{j:}$ the $j$-th row vector of $A$. We assume without loss of generality by simple scaling that[11]

$$\min_{j \in [m]} \{ \|A_{j:}\|_\infty \} = 1 \ . \tag{4.2}$$

**Proposition 4.1.** *The normalization (4.2) ensures* $\mathsf{OPT} \in [1, m]$.

*Proof.* Suppose that $j^*$ is the row that achieves the smallest infinite norm $\|A_{j:}\|_\infty$ over all rows $j \in [m]$. Then, for any solution $x \in \mathbb{R}^n_{\geq 0}$ satisfying $\langle A_{j^*:}, x \rangle \geq 1$, we must have $\mathbb{1}^T x \geq 1/\|A_{j^*:}\|_\infty = 1$ using (4.2). On the other hand, we can construct a feasible solution $x$ as follows. Initialize $x = 0$, and then for each row $j$, let us find the coordinate $i$ that maximizes the value of $A_{ij}$ among all columns $i$. Then, we increase $x_i$ by $1/A_{ij} = 1/\|A_{j:}\|_\infty$. After we have exhausted all the $m$ rows, we arrive at some $x \geq 0$ satisfying $Ax \geq \mathbb{1}$ as well as $\mathbb{1}^T x = \sum_j 1/\|A_{j:}\|_\infty \leq m$. □

In our covering LP solvers, we assume that an initial solution, achieving a constant approximation, is available to the algorithm. Such a solution can be obtained for instance by the covering LP solver from Young [37] with constant $\epsilon$ in time $O(N \log N)$.

**Definition 4.2.** *Let $x^\sharp$ be a given 2-approximate solution to the covering problem given and let* $\mathsf{OPT}' \stackrel{\text{def}}{=} \mathbb{1}^T x^\sharp \in [\mathsf{OPT}, 2\mathsf{OPT}]$. *Without loss of generality, assume* $\mathsf{OPT}' \geq 2$.

We now introduce the smoothed objective $f_\mu(x)$ we are going to minimize in order to solve covering LP. Symmetric to the case in the packing solver, this smoothed objective $f_\mu(x)$ turns each row of the LP constraint $Ax \geq \mathbb{1}$ into an exponential penalty function.

**Definition 4.3.** *Letting $\mu \stackrel{\text{def}}{=} \frac{\varepsilon}{4 \log(nm/\varepsilon)}$, we define the smoothed objective $f_\mu(x)$ as*

$$f_\mu(x) \stackrel{\text{def}}{=} \mu \sum_{j=1}^m e^{\frac{1}{\mu}(1-(Ax)_j)} + \mathbb{1}^T x \ .$$

**Fact 4.4.** $\nabla f_\mu(x) = \mathbb{1} - A^T p(x)$ *and* $\nabla^2 f_\mu(x) = \frac{1}{\mu} A^T \mathrm{diag}\{p(x)\} A$, *where* $p_j(x) \stackrel{\text{def}}{=} e^{\frac{1}{\mu}(1-(Ax)_j)}$.

We present some properties about $f_\mu(x)$. They together imply that the minimum of $f_\mu(x)$ is around $\mathsf{OPT}$, and if one approximately finds the minimum of $f_\mu(x)$ up to an additive error $O(\varepsilon\mathsf{OPT})$, this corresponds to a $(1+O(\varepsilon))$-approximate solution to the covering LP (4.1). The proofs are analogous to Section 2, and included in Appendix B for completeness' sake.

**Proposition 4.5.**
(a) $f_\mu(u^*) \leq (1+\varepsilon)\mathsf{OPT}$ *for* $u^* \stackrel{\text{def}}{=} (1+\varepsilon/2)x^*$.
(b) $f_\mu(x) \geq (1-\varepsilon)\mathsf{OPT}$ *for every* $x \geq 0$.

[11] If $\min_{j\in[m]}\{\|A_{j:}\|_\infty\} = 0$ then the covering LP is infeasible so we are done. Otherwise, if $\min_{j\in[m]}\{\|A_{j:}\|_\infty\} = v > 0$ we scale all entries of $A$ by $1/v$, and scale $\mathsf{OPT}$ by $v$.

(c) *For any $x \ge 0$ satisfying $f_\mu(x) \le 2\mathsf{OPT}$, we must have $Ax \ge (1-\varepsilon)\mathbb{1}$.*

(d) *If $x \ge 0$ satisfies $f_\mu(x) \le (1+\delta)\mathsf{OPT}$ for some $\delta \in [0,1]$, then $\frac{1}{1-\varepsilon}x$ is a $\frac{1+\delta}{1-\varepsilon}$-approximate solution to the covering LP.*

## 5 Our Covering LP Solver in the Well-Conditioned Case

Recall in packing LPs, since it satisfies $0 \le x_i^* \le \frac{1}{\|A_{:i}\|_\infty}$ (see Fact 2.1), we can minimize $f_\mu$ over a bounding box $\Delta_{\mathsf{box}}$. Unfortunately, it no longer satisfies $x_i^* \le \frac{1}{\|A_{:i}\|_\infty}$ in covering LPs, so one cannot directly turn `PacLPSolver` into its symmetric version to solve covering LP.

In this section, we show that this symmetric covering LP solver still solves all *well-behaved* covering LP instances. Specifically, we say the covering LP is well-behaved if:[12]

**Assumption 5.1.** *There exists some optimal covering LP solution $x^*$ satisfying $x_i^* \le \frac{9}{\|A_{:i}\|_\infty}$; and the initial point $x^\sharp$ satisfies $x_i^\sharp \le \frac{9}{\|A_{:i}\|_\infty}$.*

For instance, well-behaved instances naturally arise from those where the constraints $Ax \ge \mathbb{1}$ are non-redundant. If the optimal covering LP solution $x^*$ and the initial point $x^\sharp$ satisfy $\mathbb{1} \le Ax^* \le 9\cdot\mathbb{1}$ and $\mathbb{1} \le Ax^\sharp \le 9\cdot\mathbb{1}$, then Assumption 5.1 is satisfied.

Well-behaved covering LP problems immediately satisfy the following:

**Fact 5.2.** *Define $\Delta_{\mathsf{box}} \stackrel{\text{def}}{=} \{x \in \mathbb{R}^n : x_i \in \big[0, \frac{10}{\|A_{:i}\|_\infty}\big]\}$. Under Assumption 5.1, we have $u^* \stackrel{\text{def}}{=} (1+\varepsilon/2)x^* \in \Delta_{\mathsf{box}}$ and $x^{\mathsf{start}} \stackrel{\text{def}}{=} (1+\varepsilon/2)\cdot x^\sharp \in \Delta_{\mathsf{box}}$. Also, it satisfies $f_\mu(x^{\mathsf{start}}) \le 3\mathsf{OPT}$.*

*Proof.* The claims $u^*, x^{\mathsf{start}} \in \Delta_{\mathsf{box}}$ are trivial after noticing $\varepsilon \le 1/30$. Using the fact that $Ax^{\mathsf{start}} - \mathbb{1} \ge (1+\varepsilon/2)Ax^\sharp - \mathbb{1} \ge \varepsilon/2 \cdot \mathbb{1}$, we compute $f_\mu(x^{\mathsf{start}})$ as follows:

$$f_\mu(x^{\mathsf{start}}) = \mu \sum_j e^{\frac{1}{\mu}(1-(Ax^{\mathsf{start}})_j)} + \mathbb{1}^T x^{\mathsf{start}} \le \mu \sum_j e^{\frac{-\varepsilon/2}{\mu}} + 2\mathsf{OPT} \le \frac{\mu m}{(nm)^2} + 2\mathsf{OPT} < 3\mathsf{OPT} \ .$$

□

We now describe $\texttt{CovLPSolver}^{\mathsf{wb}}$ (which is a symmetric variant of `PacLPSolver`) that solves well-behaved covering LP problems, see Algorithm 2. It starts with the initial vector $\mathsf{x}_0 = \mathsf{y}_0 = x^{\mathsf{start}}$ and $\mathsf{z}_0 = 0$. Then, $\texttt{CovLPSolver}^{\mathsf{wb}}$ is divided into $T$ iterations. In each iteration $k$, it computes a weighted midpoint $\mathsf{x}_k \leftarrow \tau \mathsf{z}_{k-1} + (1-\tau)\mathsf{y}_{k-1}$ for some parameter $\tau \in (0,1)$, and then proceeds to compute $\mathsf{y}_k$ and $\mathsf{z}_k$ as follows.

We select $i \in [n]$ uniformly at random. Let $\xi_k^{(i)} = (0,\dots,0,-\mathbb{T}^{\mathsf{p}}(v),0,\dots,0)$ be the vector that is only non-zero at coordinate $i$, where $v = -\nabla_i f_\mu(\mathsf{x}_k) \in [-1,\infty)$, and recall $\mathbb{T}^{\mathsf{p}}(v) \stackrel{\text{def}}{=} \min\{v,1\}$. We refer to $\xi_k^{(i)}$ as the *truncated gradient*. Next,

- Perform a *mirror (descent) step* $\mathsf{z}_k \leftarrow \mathsf{z}_k^{(i)} \stackrel{\text{def}}{=} \arg\min_{z\in\Delta_{\mathsf{box}}} \big\{\frac{1}{2}\|z-\mathsf{z}_{k-1}\|_A^2 + \langle n\alpha_k \xi_k^{(i)}, z\rangle\big\}$ for some parameter $\alpha_k \ll 1/n$ to be chosen later.
- Perform a *gradient (descent) step* $\mathsf{y}_k \leftarrow \mathsf{y}_k^{(i)} \stackrel{\text{def}}{=} \mathsf{x}_k + \frac{1}{n\alpha_k L}(\mathsf{z}_k^{(i)} - \mathsf{z}_{k-1})$.

This finishes the description of $\texttt{CovLPSolver}^{\mathsf{wb}}$. It is not surprising to deduce the following theorem similar to Theorem 3.4. We include its proof in Appendix C for completeness.

[12] The constant 9 in this section can be replaced with any other constant greater than 1.

**Algorithm 2** $\texttt{CovLPSolver}^{\mathsf{wb}}(A, x^{\mathsf{start}}, \varepsilon)$

**Input:** $A \in \mathbb{R}_{\geq 0}^{m \times n}, x^{\mathsf{start}} \in \Delta_{\mathsf{box}}, \varepsilon \in (0, 1/30]$.

**Output:** $x \in \Delta_{\mathsf{box}}$. ▷ recall $\Delta_{\mathsf{box}} \stackrel{\text{def}}{=} \{x \in \mathbb{R}^n : x_i \in \big[0, \frac{10}{\|A_{:i}\|_\infty}\big]\}$

$\mu \leftarrow \frac{\varepsilon}{4\log(nm/\varepsilon)}$, $L \leftarrow \frac{4}{\mu}$, $\tau \leftarrow \frac{1}{21 \cdot nL}$ and $\alpha_0 \leftarrow \frac{1}{nL}$. ▷ parameters

$T \leftarrow \lceil 21 nL \log(1/\varepsilon) \rceil = O(n \cdot \frac{\log(nm/\varepsilon) \cdot \log(1/\varepsilon)}{\varepsilon})$. ▷ number of iterations

$\mathsf{x}_0 = \mathsf{y}_0 \leftarrow x^{\mathsf{start}}$, $\mathsf{z}_0 \leftarrow 0$.

**for** $k \leftarrow 1$ **to** $T$ **do**

  $\alpha_k \leftarrow \frac{1}{1-\tau} \alpha_{k-1}$

  $\mathsf{x}_k \leftarrow \tau \mathsf{z}_{k-1} + (1-\tau)\mathsf{y}_{k-1}$.

  Randomly select $i \in [n]$ uniformly at random.

  Define vector $\xi_k^{(i)}$ be all-zero except at coordinate $i$, where $\xi_{k,i}^{(i)} = \max\{-1, \nabla_i f_\mu(\mathsf{x}_k)\}$.

  $\mathsf{z}_k \leftarrow \mathsf{z}_k^{(i)} \stackrel{\text{def}}{=} \arg\min_{z \in \Delta_{\mathsf{box}}} \big\{\frac{1}{2}\|z - \mathsf{z}_{k-1}\|_A^2 + \langle n\alpha_k \xi_k^{(i)}, z\rangle\big\}$. ▷ See Proposition 3.2

  $\mathsf{y}_k \leftarrow \mathsf{y}_k^{(i)} \stackrel{\text{def}}{=} \mathsf{x}_k + \frac{1}{n\alpha_k L}(\mathsf{z}_k^{(i)} - \mathsf{z}_{k-1})$.

**end for**

**return** $\mathsf{y}_T$.

**Theorem 5.3.** *Under the well-behavior assumption 5.1 in the covering LP problem,* $\texttt{CovLPSolver}^{\mathsf{wb}}(A, x^{\mathsf{start}}, \varepsilon)$ *outputs some* $\mathsf{y}_T$ *satisfying*

$$\mathbf{E}[f_\mu(\mathsf{y}_T)] \leq (1 + 4.6\varepsilon)\mathsf{OPT} \enspace .$$

Again, using the same proof as Corollary 3.5, one can apply Markov's bound to turn Theorem 5.3 into a probabilistic statement:

**Corollary 5.4.** *Under the well-behavior assumption 5.1 in the covering LP problem, with probability at least 2/3,* $\mathsf{y}_T = \texttt{CovLPSolver}^{\mathsf{wb}}(A, x^{\mathsf{start}}, \varepsilon)$ *satisfies that* $\frac{\mathsf{y}_T}{1-\varepsilon}$ *is a* $(1 + O(\varepsilon))$ *approximate solution to covering LP. The expected running time is*

$$O\Big(\frac{\log(nm/\varepsilon)\log(1/\varepsilon)}{\varepsilon} N\Big) \enspace .$$

**Removing the well-behavior assumption.** In subsequent work, after the conference presentation of this paper, Wang, Mahoney and Rao [35] showed the following theorem.

**Theorem 5.5** (Wang et al. [35])**.** *Any covering LP with constraint matrix* $A \in \mathbb{R}_{\geq 0}^{m \times n}$ *of sparsity* $N$ *can be converted into an equivalent but* well-behaved *covering LP with matrix* $\widetilde{A} \in \mathbb{R}^{m \times n \cdot O(\log(mn/\varepsilon))}$ *and sparsity* $N \cdot O(\log(mn/\varepsilon))$*. The conversion takes time* $N \cdot O(\log(mn/\varepsilon))$*.*

As a result, we can apply our covering solver $\texttt{CovLPSolver}^{\mathsf{wb}}$ to this modified LP and apply our Theorem 5.3 to solve any covering LP in expected time $O(\frac{\log^2(nm/\varepsilon)\log(1/\varepsilon)}{\varepsilon} N)$.

# 6 Our Covering LP Solver in the General Case

In this section, we remove the well-behavior assumption and propose a different algorithm $\texttt{CovLPSolver}$ to solve all covering LP instances. This algorithm introduces a factor $1/\sqrt{\varepsilon}$ loss in the running time, but is a *direct* covering LP solver without using any reduction.

The main difference to $\texttt{PacLPSolver}$ and $\texttt{CovLPSolver}^{\mathsf{wb}}$ is that, this time we abandon the box

**Algorithm 3** $\texttt{CovLPSolver}(A, x^{\mathsf{start}}, \varepsilon)$

**Input:** $A \in \mathbb{R}_{\geq 0}^{m \times n}$, $x^{\mathsf{start}} \in \Delta_{\mathsf{simplex}}$, $\varepsilon \in (0, 1/30]$.
**Output:** $x \in \Delta_{\mathsf{simplex}}$.

$\mu \leftarrow \frac{\varepsilon}{4\log(nm/\varepsilon)}$, $\beta \leftarrow \sqrt{\varepsilon}$, $\tau \leftarrow \frac{\mu\beta}{12n}$. ▷ parameters
$T \leftarrow \lceil \frac{1}{\tau}\log(1/\varepsilon) \rceil = O(\frac{\log(nm/\varepsilon)\log(1/\varepsilon)}{\varepsilon^{1.5}} n)$. ▷ number of iterations
$\alpha_0 \leftarrow (1-\tau)^T \frac{\varepsilon}{12n\beta}$ and $\gamma \leftarrow \frac{\varepsilon}{6\beta}$. ▷ so that $\alpha_T = \frac{\varepsilon}{12n\beta}$ and $\gamma = 2\alpha_T n$
$\mathsf{x}_0 = \mathsf{y}_0 = \mathsf{z}_0 \leftarrow x^{\mathsf{start}}$.
**for** $k \leftarrow 1$ **to** $T$ **do**
  $\alpha_k \leftarrow \frac{1}{1-\tau}\alpha_{k-1}$.
  $\mathsf{x}_k \leftarrow \tau \mathsf{z}_{k-1} + (1-\tau)\mathsf{y}_{k-1}$.
  Randomly select $i$ uniformly at random from $[n]$.
  Define vector $\xi_k^{(i)}$ to be all-zero except at coordinate $i$, where $\xi_{k,i}^{(i)} = \max\{-\beta, \nabla_i f_\mu(\mathsf{x}_k)\}$.
  $\mathsf{z}_k \leftarrow \mathsf{z}_k^{(i)} \stackrel{\text{def}}{=} \arg\min_{z \in \Delta_{\mathsf{simplex}}} \big\{ V_{\mathsf{z}_{k-1}}(z) + \langle (1+\gamma) n \alpha_k \xi_k^{(i)}, z \rangle \big\}$. ▷ See Proposition 6.4
  **if** $\nabla_i f_\mu(\mathsf{x}_k) < -\beta$ **then**
    Denote by $\pi$ the permutation that satisfies $A_{\pi(1),i} \leq \cdots \leq A_{\pi(m),i}$.
    Pick $j^* \in [m]$ such that $\begin{cases} \sum_{j<j^*} A_{\pi(j),i} \cdot p_{\pi(j)}(\mathsf{x}_k) < 1+\beta \\ \sum_{j \leq j^*} A_{\pi(j),i} \cdot p_{\pi(j)}(\mathsf{x}_k) \geq 1+\beta \end{cases}$
    ▷ $j^* \in [m]$ always exists, see (6.1)
    $\mathsf{y}_k \leftarrow \mathsf{y}_k^{(i)} \stackrel{\text{def}}{=} \mathsf{x}_k + \delta \cdot \mathbf{e}_i$ where $\delta = \frac{\mu\beta}{2A_{\pi(j^*),i}}$.
  **else**
    $\mathsf{y}_k \leftarrow \mathsf{y}_k^{(i)} \stackrel{\text{def}}{=} \mathsf{x}_k$.
  **end if**
**end for**
**return** $\mathsf{y}_T$.

constraint and study the minimization of $f_\mu(x)$ over a simplex

$$x \in \Delta_{\mathsf{simplex}} \stackrel{\text{def}}{=} \{x \in \mathbb{R}^n \,:\, x_i \geq 0 \,\wedge\, \mathbb{1}^T x \leq 2\mathsf{OPT}'\} \ .$$

Again, this constraint $\mathbb{1}^T x \leq 2\mathsf{OPT}'$ is redundant just like the old $\Delta_{\mathsf{box}}$ constraint for packing LP (recall Remark 2.5); however, it shall be used to make sure that our updates are always inside $\Delta_{\mathsf{simplex}}$. It is a simple fact that

**Fact 6.1.** $u^* \stackrel{\text{def}}{=} (1+\varepsilon/2)x^* \in \Delta_{\mathsf{simplex}}$.

Recall that the initial vector $x^\sharp$ is defined in Definition 4.2, and $\mathsf{OPT}'$ is a crude approximation to $\mathsf{OPT}$, satisfying $\mathsf{OPT}' \stackrel{\text{def}}{=} \mathbb{1}^T x^\sharp \in [\mathsf{OPT}, 2\mathsf{OPT}]$. We choose different starting vector $x^{\mathsf{start}}$ from Section 5:

**Proposition 6.2.** *Letting* $x^{\mathsf{start}} \stackrel{\text{def}}{=} (1+\varepsilon/2) \cdot x^\sharp + (\frac{1}{n}, \dots, \frac{1}{n})$, *we have* $x^{\mathsf{start}} \in \Delta_{\mathsf{simplex}}$ *and* $f_\mu(x^{\mathsf{start}}) \leq 4\mathsf{OPT}$.

*Proof.* Using $Ax^{\mathsf{start}} - \mathbb{1} \geq (1+\varepsilon/2)Ax^\sharp - \mathbb{1} \geq \varepsilon/2 \cdot \mathbb{1}$, we compute $f_\mu(x^{\mathsf{start}})$ as follows:

$$f_\mu(x^{\mathsf{start}}) = \mu \sum_j e^{\frac{1}{\mu}(1-(Ax^{\mathsf{start}})_j)} + \mathbb{1}^T x^{\mathsf{start}} \leq \mu \sum_j e^{\frac{-\varepsilon/2}{\mu}} + 2\mathsf{OPT} + 1 \leq \frac{\mu m}{(nm)^2} + 3\mathsf{OPT} < 4\mathsf{OPT} \ .$$

Also, we have $\mathbb{1}^T x^{\mathsf{start}} \leq (1+\varepsilon/2)\mathsf{OPT}' + 1 \leq 2\mathsf{OPT}'$. (Recall $\mathsf{OPT}' \geq 2$ in Definition 4.2.) □

Our proposed algorithm $\texttt{CovLPSolver}$ starts with the initial vector $\mathsf{x}_0 = \mathsf{y}_0 = \mathsf{z}_0 = x^{\mathsf{start}}$ and is divided into $T$ iterations. In each iteration $k$, as usual, it computes a weighted midpoint $\mathsf{x}_k \leftarrow \tau \mathsf{z}_{k-1} + (1-\tau)\mathsf{y}_{k-1}$ for some parameter $\tau \in (0,1)$, and then computes $\mathsf{y}_k$ and $\mathsf{z}_k$ as follows.

We select $i \in [n]$ uniformly at random, and let $\xi_k^{(i)} = (0,\dots,0,\mathbb{T}^{\mathsf{c}}(v),0,\dots,0)$ be the vector that is only non-zero at coordinate $i$, where $v = \nabla_i f_\mu(\mathsf{x}_k) \in (-\infty, 1]$ and $\mathbb{T}^{\mathsf{c}}(v) \stackrel{\text{def}}{=} \max\{-\beta, v\}$ is the new thresholding function for some parameter $\beta \stackrel{\text{def}}{=} \sqrt{\varepsilon}$. Then,

- Perform a *mirror (descent) step* $\mathsf{z}_k \leftarrow \mathsf{z}_k^{(i)} \stackrel{\text{def}}{=} \arg\min_{z \in \Delta_{\mathsf{simplex}}} \big\{ V_{\mathsf{z}_{k-1}}(z) + \langle (1+\gamma) n \alpha_k \xi_k^{(i)}, z \rangle \big\}$ for some positive parameters $\gamma \ll 1$ and $\alpha_k \ll 1/n$, where
$$V_x(y) \stackrel{\text{def}}{=} \textstyle\sum_{i=1}^n y_i \log \frac{y_i}{x_i} + x_i - y_i$$
is the so-called Bregman divergence of the generalized entropy function (c.f. [13]).
- If $\nabla_i f_\mu(x_k) < -\beta$, perform a *gradient (descent) step* $\mathsf{y}_k \leftarrow \mathsf{y}_k^{(i)} \stackrel{\text{def}}{=} \mathsf{x}_k + \delta \mathbf{e}_i$ for some $\delta > 0$. In practice, one can line-search over $\delta$, but we choose an explicit $\delta$ as follows.
  - Denote by $\pi$ the permutation that satisfies $A_{\pi(1),i} \le \cdots \le A_{\pi(m),i}$
  - Pick $j^* \in [m]$ s.t. $\begin{cases} \sum_{j<j^*} A_{\pi(j),i} \cdot p_{\pi(j)}(\mathsf{x}_k) < 1+\beta \\ \sum_{j\le j^*} A_{\pi(j),i} \cdot p_{\pi(j)}(\mathsf{x}_k) \ge 1+\beta \end{cases}$. Such $j^*$ exists because
$$\textstyle\sum_{j=1}^m A_{\pi(j),i} \cdot p_{\pi(j)}(\mathsf{x}_k) = \sum_{j=1}^m A_{ji} \cdot p_j(\mathsf{x}_k) = 1 - \nabla_i f_\mu(\mathsf{x}_k) \ge 1+\beta \ . \qquad (6.1)$$
  - Set $\delta = \frac{\mu\beta}{2A_{\pi(j^*),i}}$.

This finishes the description of our `CovLPSolver`.

**Remark 6.3.** We use the superscript $^{(i)}$ on $\xi_k^{(i)}$, $\mathsf{y}_k^{(i)}$ and $\mathsf{z}_k^{(i)}$ to emphasize that the value depends on the choice of $i$. We have used generic parameters $\tau, \alpha_k, T$ in the above description and their precise values are presented in Algorithm 3.

Our update on $\mathsf{y}_k$ is a "gradient descent step" because we shall prove that it strictly decreases the objective (i.e., $f_\mu(\mathsf{x}_k) - f_\mu(\mathsf{y}_k^{(i)}) \ge 0$). Our update on $\mathsf{z}_k$ is a "mirror descent step" because we shall apply standard mirror descent analysis [13] to it. We explicitly describe how to implement this mirror step: (proved in Appendix D)

**Proposition 6.4.** *If $\mathsf{z}_{k-1} \in \Delta_{\mathsf{simplex}}$ and $\mathsf{z}_{k-1} > 0$, the minimizer $z = \arg\min_{z \in \Delta_{\mathsf{simplex}}} \big\{ V_{\mathsf{z}_{k-1}}(z) + \langle \delta \mathbf{e}_i, z \rangle \big\}$ for any scalar $\delta \in \mathbb{R}$ and basis vector $\mathbf{e}_i$ can be computed as follows:*
1. *$z \leftarrow \mathsf{z}_{k-1}$.*
2. *$z_i \leftarrow z_i \cdot e^{-\delta}$.*
3. *If $\mathbb{1}^T z > 2\mathsf{OPT}'$, $z \leftarrow \frac{2\mathsf{OPT}'}{\mathbb{1}^T z} z$.*
4. *Return $z$.*

We also point out that

**Lemma 6.5.** *Each iteration of* `CovLPSolver` *can be implemented to run in expected $O(N/n)$ time. The total expected running time is $O(TN/n)$.*

The proof of Lemma 6.5 is analogous to its packing counterpart, and included in Section F.

## 6.1 Main Proof Ideas and Ingredients

In `CovLPSolver`, we pick a random coordinate $i \in [n]$ at each iteration, and decompose $\nabla_i f(\mathsf{x}_k) = \xi + \eta$, where $\eta \in (-\infty, 0]$ is the (negative) large gradient component and $\xi \in [-\sqrt{\varepsilon}, 1]$ is the truncated gradient component. In other words, we truncate the gradient $\nabla_i f(\mathsf{x}_k)$ at a negative threshold $-\beta = -\sqrt{\varepsilon}$, rather than at $-1$ as in $\texttt{CovLPSolver}^{\mathsf{wb}}$.

The reason for this new threshold $-\sqrt{\varepsilon}$ can be understood as follows. In PacLPSolver (and symmetrically in $\texttt{CovLPSolver}^{\mathsf{wb}}$), we used Lemma 3.10 to show that our gradient descent step $\mathsf{y}_k$ decreases the objective by an amount that both includes the $\xi$ and $\eta$ components. Unfortunately, for covering LP, this decrease amount is only proportional to $\eta$ but not to $\xi$ (compare Lemma 3.10 with Lemma 6.14 later). This forces us to treat the small gradient $\xi$ separately using mirror descent, but not gradient descent.

If $\xi$ were in $[-1, 1]$, classical theory of mirror descent [13] (or multiplicative weight update [8]) would imply that the mirror step $\mathsf{z}_k$ converges at a rate $\propto \varepsilon^{-2}$. This is too slow. Instead, since truncated $\xi$ into a smaller interval $[-\sqrt{\varepsilon}, 1]$, using a *negative-width technique* (see Section 6.5), we improve this mirror-descent convergence rate from $\varepsilon^{-2}$ to $\varepsilon^{-1.5}$.

On the other hand, due to this truncation at $-\sqrt{\varepsilon}$ instead of $-1$, our gradient step on $\mathsf{y}_k$ also converges slower, at a rate $1/\varepsilon^{1.5}$ instead of $1/\varepsilon$. This is why $\beta = \sqrt{\varepsilon}$ is the best truncation threshold, as it balances gradient and mirror descent.

Another ingredient behind our proof is a new distance bound that is uncommon in first-order analysis. Recall that, given convex function $g(x)$, traditional analysis applies convexity argument $g(x) - g(x^*) \leq \langle \nabla g(x), x - x^* \rangle$ to bound the objective distance to optimum. If $g(x) = e^{-x}$ is univariate, $x = -1$, and $x^* = -100$, this bound becomes $e^{-1} \approx e^{-1} - e^{-100} \leq e^{-1} \cdot 99$, which is very loose. In our analysis, we replace this convexity argument with a more benign bound, specifically designed for covering LP (see Lemma 6.10).

## 6.2 Convergence Statement

The main convergence theorem of this section is as follows:

**Theorem 6.6.** $\texttt{CovLPSolver}(A, x^{\mathsf{start}}, \varepsilon)$ *outputs some* $\mathsf{y}_T$ *satisfying*
$$\mathbf{E}[f_\mu(\mathsf{y}_T)] \leq (1 + 9\varepsilon)\mathsf{OPT} \ .$$

Again, using the same proof as Corollary 3.5, one can apply Markov's bound to turn Theorem 5.3 into a probabilistic statement:

**Corollary 6.7.** *With probability at least* $2/3$, $\mathsf{y}_T = \texttt{CovLPSolver}(A, x^{\mathsf{start}}, \varepsilon)$ *satisfies that* $\frac{\mathsf{y}_T}{1-\varepsilon}$ *is a* $(1 + O(\varepsilon))$ *approximate solution to the covering LP program. The expected running time is* $O(\frac{\log(nm/\varepsilon)\log(1/\varepsilon)}{\varepsilon^{1.5}} N)$.

Before delving into the proof of Theorem 6.6, we make the following observations:

**Fact 6.8.** *For every* $k \in \{0, 1, \dots, T\}$, *it satisfies* $\mathsf{x}_k, \mathsf{y}_k \geq 0$, $\mathsf{z}_k > 0$, *and* $\mathsf{z}_k \in \Delta_{\mathsf{simplex}}$ .

*Proof.* Since the $x^{\mathsf{start}}$ satisfies $\mathbb{1}^T x^{\mathsf{start}} \leq 2\mathsf{OPT}'$ by Proposition 6.2, we have $\mathsf{z}_0 = x^{\mathsf{start}} \in \Delta_{\mathsf{simplex}}$. Also, the mirror descent step (see Proposition 6.4) ensures $\mathsf{z}_{k,i} > 0$ for all rounds $k$ and coordinates $i$, as well as $\mathsf{z}_k \in \Delta_{\mathsf{simplex}}$ for all rounds $k$. However, we $\mathsf{x}_k$ and $\mathsf{y}_k$ may not necessarily lie inside $\Delta_{\mathsf{simplex}}$, but will always stay non-negative. □

We prove Theorem 6.6 in the subsequent subsections.

## 6.3 Step 1: Distance Adjustment

Using convexity, one can obtain
$$f_\mu(\mathsf{x}_k) - f_\mu(u) \leq \langle \nabla f_\mu(\mathsf{x}_k), \mathsf{x}_k - u \rangle \quad \text{for every } u \in \Delta_{\mathsf{simplex}}. \tag{6.2}$$

Note that inequality (6.2) can be very loose for exponential functions. For instance, if $f_\mu(x)$ were as simple as $e^x$, then the convexity inequality $e^b - e^a \le e^b \cdot (b-a)$ says

- when $b = 2$ and $a = -10$, we have $e^2 - e^{-10} \le 12e^2$;
- when $b = 2$ and $a = -100$, we have $e^2 - e^{-100} \le 102e^2$.

Although $e^{-100} \approx e^{-10}$, the two upper bounds are off from each other by a factor of 10.

In this section, we strengthen (6.2) in the special case of $u = u^* \stackrel{\text{def}}{=} (1+\varepsilon/2)x^*$. For analysis purpose, let $\widetilde{A}$ be the *adjusted matrix* of $A$ described as follows.

**Definition 6.9** (adjusted matrix $\widetilde{A}$)**.** *For each row $j \in [m]$, if $(Au^*)_j \le 2$ then we keep it and let $\widetilde{A}_{j:} \stackrel{\text{def}}{=} A_{j:}$. Otherwise, —that is, if $(Au^*)_j > 2$— we define $\widetilde{A}_{j:} \stackrel{\text{def}}{=} \frac{2}{(Au^*)_j} \cdot A_{j:}$ to be the same $j$-th row $A_{j:}$, but scaled down by a factor of $\frac{2}{(Au^*)_j}$. It is clear from this definition that*

$$A_{ji} \ge \widetilde{A}_{ji} \text{ for all } (i,j) \in [n] \times [m] \quad \text{and} \quad (1+\varepsilon)\mathbb{1} \le \widetilde{A}u^* \le 2\mathbb{1} .$$

**Lemma 6.10** (distance adjustment)**.**

$$\begin{aligned} f_\mu(\mathsf{x}_k) - f_\mu(u^*) &\le \langle \mathbb{1} - A^T p(\mathsf{x}_k), \mathsf{x}_k - u^* \rangle + \langle \widetilde{A}^T p(\mathsf{x}_k) - A^T p(\mathsf{x}_k), u^* \rangle + \varepsilon \mathsf{OPT} \\ &= \langle \nabla f_\mu(\mathsf{x}_k), \mathsf{x}_k - u^* \rangle + \langle \widetilde{A}^T p(\mathsf{x}_k) - A^T p(\mathsf{x}_k), u^* \rangle + \varepsilon \mathsf{OPT} \end{aligned}$$

At high level, ignoring the negligible term $\varepsilon\mathsf{OPT}$, Lemma 6.10 strengthens the classical bound due to the extra term of $\langle \widetilde{A}^T p(\mathsf{x}_k) - A^T p(\mathsf{x}_k), u^* \rangle$. This extra term is always non-positive since $\widetilde{A} \le A$ coordinate-wise, but may be very negative in certain cases.

*Proof of Lemma 6.10.*

$$\begin{aligned} f_\mu(\mathsf{x}_k) - f_\mu(u^*) &= \mu \sum_{j=1}^m \Big( e^{\frac{1}{\mu}(1-(A\mathsf{x}_k)_j)} - e^{\frac{1}{\mu}(1-(Au^*)_j)} \Big) + \langle \mathbb{1}, \mathsf{x}_k - u^* \rangle \\ &\overset{①}{\le} \mu \sum_{j=1}^m \Big( e^{\frac{1}{\mu}(1-(A\mathsf{x}_k)_j)} - e^{\frac{1}{\mu}(1-(\widetilde{A}u^*)_j)} \Big) + \langle \mathbb{1}, \mathsf{x}_k - u^* \rangle + \mu \cdot m \cdot e^{-1/\mu} \\ &\overset{②}{\le} \sum_{j=1}^m e^{\frac{1}{\mu}(1-(A\mathsf{x}_k)_j)} \cdot \big( (\widetilde{A}u^*)_j - (A\mathsf{x}_k)_j \big) + \langle \mathbb{1}, \mathsf{x}_k - u^* \rangle + \varepsilon \mathsf{OPT} \\ &= \sum_{j=1}^m p_j(\mathsf{x}_k) \cdot \big( (\widetilde{A}u^*)_j - (A\mathsf{x}_k)_j \big) + \langle \mathbb{1}, \mathsf{x}_k - u^* \rangle + \varepsilon \mathsf{OPT} \\ &= \sum_{j=1}^m p_j(\mathsf{x}_k) \cdot \big( (Au^*)_j - (A\mathsf{x}_k)_j \big) + \langle \mathbb{1}, \mathsf{x}_k - u^* \rangle + \sum_{j=1}^m p_j(\mathsf{x}_k) \cdot \big( (\widetilde{A}u^*)_j - (Au^*)_j \big) + \varepsilon \mathsf{OPT} \\ &= \langle -A^T p(\mathsf{x}_k), \mathsf{x}_k - u^* \rangle + \langle \mathbb{1}, \mathsf{x}_k - u^* \rangle + \langle \widetilde{A}^T p(\mathsf{x}_k) - A^T p(\mathsf{x}_k), u^* \rangle + \varepsilon \mathsf{OPT} \ . \end{aligned}$$

Above, ① is because if $(Au^*)_j \ne (\widetilde{A}u^*)_j$ for some $j$, then it must satisfy that $(\widetilde{A}u^*)_j = 2$, and therefore $-e^{\frac{1}{\mu}(1-(Au^*)_j)} \le -e^{\frac{1}{\mu}(1-(\widetilde{A}u^*)_j)} + e^{-1/\mu}$. ② uses the convexity inequality of $e^b - e^a \le e^b \cdot (b-a)$, and the fact that $\mu m e^{-1/\mu} \ll \varepsilon\mathsf{OPT}$. □

## 6.4 Step 2: Gradient Truncation

For analysis purpose, let us separate the indices $i \in [n]$ into large and small ones.

**Definition 6.11.** *We make the following definitions.*

- *Let* $B_k \stackrel{\text{def}}{=} \{i \in [n] : \nabla_i f_\mu(\mathsf{x}_k) < -\beta\}$ *and* $[n] \setminus B_k$ *be the set of* large and small indices.
- *Let* $\xi_k \in [-\beta, 1]^n$ *be the* truncated gradient *so that* $\xi_{k,i} = \mathbb{T}^{\mathsf{c}}(\nabla_i f_\mu(x_k))$ *for each* $i \in [n]$.
- *Let* $\eta_k \in (-\infty, 0]^n$ *be the* large gradient *so that* $\nabla f_\mu(\mathsf{x}_k) = \xi_k + \eta_k$. *It is clear that*

  $$\eta_{k,i} = 0 \text{ for every } i \notin B, \text{ and } \eta_{k,i} = (1+\beta) - (A^T p(\mathsf{x}_k))_i \text{ for every } i \in B.$$
- *Let* $\widetilde{\eta}_k \in (-\infty, \infty)^n$ *be the* adjusted large gradient *so that*

  $$\widetilde{\eta}_{k,i} = 0 \text{ for every } i \notin B, \text{ and } \widetilde{\eta}_{k,i} = (1+\beta) - (\widetilde{A}^T p(\mathsf{x}_k))_i \text{ for every } i \in B.$$

*We denote by* $\eta_k^{(i)} = (0, \dots, 0, \eta_{k,i}, 0, \dots, 0)$, *the vector that is zero at all coordinates other than* $i$, *and similarly* $\xi_k^{(i)} = (0, \dots, \xi_{k,i}, \dots, 0)$ *and* $\widetilde{\eta}_k^{(i)} = (0, \dots, \widetilde{\eta}_{k,i}, \dots, 0)$. *We emphasize that* $\eta_k^{(i)} \neq \eta_k$, $\widetilde{\eta}_k^{(i)} \neq \widetilde{\eta}_k$, *and* $\xi_k^{(i)} \neq \xi_k$.

The following key lemma is very analogous to (3.1) in the packing LP analysis.

**Lemma 6.12** (distance upper bound)**.**

$$f_\mu(\mathsf{x}_k) - f_\mu(u^*) \le \frac{(1-\tau)}{\tau}(f_\mu(\mathsf{y}_{k-1}) - f_\mu(\mathsf{x}_k)) + \mathbf{E}_i\Big[\langle n\xi_k^{(i)}, \mathsf{z}_{k-1} - u^*\rangle\Big] + \mathbf{E}_i\Big[\langle n\widetilde{\eta}_k^{(i)}, -u^*\rangle\Big] + \varepsilon\mathsf{OPT} \enspace .$$

Note that if one uses $\eta_k^{(i)}$ instead of $\widetilde{\eta}_k^{(i)}$, then Lemma 6.12 becomes trivial to prove just like (3.1). The reason we can have the stronger term $\widetilde{\eta}_k^{(i)}$ is precisely due to the distance adjustment Lemma 6.10.

*Proof of Lemma 6.12.* We derive the following sequence of inequalities:

$$\begin{aligned}
&\big(f_\mu(\mathsf{x}_k) - f_\mu(u^*)\big) - \varepsilon\mathsf{OPT} \\
&\stackrel{①}{\le} \langle \nabla f_\mu(\mathsf{x}_k), \mathsf{x}_k - u^*\rangle + \langle \widetilde{A}^T p(\mathsf{x}_k) - A^T p(\mathsf{x}_k), u^*\rangle \\
&= \langle \nabla f_\mu(\mathsf{x}_k), \mathsf{x}_k - \mathsf{z}_{k-1}\rangle + \langle \nabla f_\mu(\mathsf{x}_k), \mathsf{z}_{k-1} - u^*\rangle + \langle \widetilde{A}^T p(\mathsf{x}_k) - A^T p(\mathsf{x}_k), u^*\rangle \\
&\stackrel{②}{=} \frac{(1-\tau)}{\tau}\langle \nabla f_\mu(\mathsf{x}_k), \mathsf{y}_{k-1} - \mathsf{x}_k\rangle + \langle \nabla f_\mu(\mathsf{x}_k), \mathsf{z}_{k-1} - u^*\rangle + \langle \widetilde{A}^T p(\mathsf{x}_k) - A^T p(\mathsf{x}_k), u^*\rangle \\
&\stackrel{③}{\le} \frac{(1-\tau)}{\tau}(f_\mu(\mathsf{y}_{k-1}) - f_\mu(\mathsf{x}_k)) + \langle \nabla f_\mu(\mathsf{x}_k), \mathsf{z}_{k-1} - u^*\rangle + \langle \widetilde{A}^T p(\mathsf{x}_k) - A^T p(\mathsf{x}_k), u^*\rangle \\
&= \frac{(1-\tau)}{\tau}(f_\mu(\mathsf{y}_{k-1}) - f_\mu(\mathsf{x}_k)) + \langle \xi_k + \eta_k, \mathsf{z}_{k-1} - u^*\rangle + \langle \widetilde{A}^T p(\mathsf{x}_k) - A^T p(\mathsf{x}_k), u^*\rangle \\
&\stackrel{④}{\le} \frac{(1-\tau)}{\tau}(f_\mu(\mathsf{y}_{k-1}) - f_\mu(\mathsf{x}_k)) + \langle \xi_k, \mathsf{z}_{k-1} - u^*\rangle + \langle \widetilde{A}^T p(\mathsf{x}_k) - A^T p(\mathsf{x}_k) - \eta_k, u^*\rangle \\
&\stackrel{⑤}{\le} \frac{(1-\tau)}{\tau}(f_\mu(\mathsf{y}_{k-1}) - f_\mu(\mathsf{x}_k)) + \langle \xi_k, \mathsf{z}_{k-1} - u^*\rangle + \langle -\widetilde{\eta}_k, u^*\rangle \\
&= \frac{(1-\tau)}{\tau}(f_\mu(\mathsf{y}_{k-1}) - f_\mu(\mathsf{x}_k)) + \mathbf{E}_i\Big[\langle n\xi_k^{(i)}, \mathsf{z}_{k-1} - u^*\rangle + \langle -n\widetilde{\eta}_k^{(i)}, u^*\rangle\Big] \enspace .
\end{aligned}$$

Above, ① is due to Lemma 6.10. ② is because $\mathsf{x}_k = \tau \mathsf{z}_{k-1} + (1-\tau)\mathsf{y}_{k-1}$, which implies that $\tau(\mathsf{x}_k - \mathsf{z}_{k-1}) = (1-\tau)(\mathsf{y}_{k-1} - \mathsf{x}_k)$. ③ is by the convexity of $f_\mu(\cdot)$. ④ is because $\langle \eta_k, \mathsf{z}_{k-1}\rangle \le 0$, since $\eta_k \le 0$ while $\mathsf{z}_{k-1} \ge 0$. ⑤ needs some careful justification: for every $i \notin B_k$, we have $(\widetilde{A}^T p(\mathsf{x}_k) - A^T p(\mathsf{x}_k))_i - \eta_{k,i} \le 0 - 0 = -\widetilde{\eta}_{k,i}$; for every $i \in B_k$, we have

$$\begin{aligned}
(\widetilde{A}^T p(\mathsf{x}_k) - A^T p(\mathsf{x}_k))_i - \eta_{k,i} &= (\widetilde{A}^T p(\mathsf{x}_k) - A^T p(\mathsf{x}_k))_i - \big((1+\beta) - (A^T p(\mathsf{x}_k))_i\big) \\
&= -\big((1+\beta) - (\widetilde{A}^T p(\mathsf{x}_k))_i\big) = -\widetilde{\eta}_{k,i} \enspace ,
\end{aligned}$$

where the two equalities follow from the definitions of $\eta_{k,i}$ and $\widetilde{\eta}_{k,i}$. □

### 6.5 Step 3: Mirror Descent Guarantee

Our update $\mathsf{z}_k^{(i)} \stackrel{\mathrm{def}}{=} \arg\min_{z\in\Delta_{\mathsf{simplex}}} \big\{V_{\mathsf{z}_{k-1}}(z) + \langle (1+\gamma)n\alpha_k\xi_k^{(i)}, z\rangle\big\}$ is, by its definition, a mirror descent step [13]. We begin by explaining an attempt that is too weak for obtaining the $\varepsilon^{-1.5}$ convergence rate.

Using the classical theory, it is not hard to repeat the proof of Lemma 3.7 —although changing the distance function from $\|\cdot\|_A^2$ to $V_x(y)$— and obtain that, as long as $\xi_{k,i}$ is in $[-1,+1]$ for each coordinate $i$, for every $u \in \Delta_{\mathsf{simplex}}$,

$$\mathbf{E}_i\big[\alpha_k\langle n\xi_k^{(i)}, \mathsf{z}_{k-1}-u\rangle\big] \leq V_{\mathsf{z}_{k-1}}(u) - \mathbf{E}_i\Big[V_{\mathsf{z}_k^{(i)}}(u)\Big] + O(\alpha_k^2 n)\mathsf{OPT} \enspace .$$

This inequality only yields a slower $\varepsilon^{-2}$ convergence rate, and $\pm 1$ is also know as the *width parameter* from the multiplicative-weight-update language [8].

In our lemma below, we make use of the fact $\xi_{k,i}$ is in $[-\beta,+1] \subseteq [-1,+1]$. In essence, this allows us to replace the $O(\alpha_k^2 n)$ factor with a better $O(\alpha_k^2\beta n)$ factor. We call it the *negative-width technique.*[13] Formally,

**Lemma 6.13** (mirror descent)**.** *Denoting by $\gamma \stackrel{\mathrm{def}}{=} 2\alpha_T n$, we have*

$$\mathbf{E}_i\big[\alpha_k\langle n\xi_k^{(i)}, \mathsf{z}_{k-1}-u^*\rangle\big] \leq V_{\mathsf{z}_{k-1}}\big(\frac{u^*}{1+\gamma}\big) - \mathbf{E}_i\Big[V_{\mathsf{z}_k^{(i)}}\big(\frac{u^*}{1+\gamma}\big)\Big] + 12\mathsf{OPT}\cdot\gamma\alpha_k\beta \enspace .$$

The proof is somewhat technical and included in Appendix D.

### 6.6 Step 4: Gradient Descent Guarantee

We show our gradient step never increases the objective for all choices of $i$. In addition, it decreases the objective by an amount proportional to the adjusted large gradient $\widetilde{\eta}_k^{(i)}$.

**Lemma 6.14** (gradient descent)**.** *For every $i \in [n]$, we have*

*(a)* $f_\mu(\mathsf{x}_k) - f_\mu(\mathsf{y}_k^{(i)}) \geq 0$, *and*

*(b)* $f_\mu(\mathsf{x}_k) - f_\mu(\mathsf{y}_k^{(i)}) \geq \frac{\mu\beta}{12}\cdot\langle -\widetilde{\eta}_k^{(i)}, u^*\rangle$ .

The proof of Lemma 6.14 is quite technical and can be found in Appendix D.

At high level, one would hope to prove that the gradient step decreases the objective by an amount proportional to the large gradient $\eta_k^{(i)}$, rather than the adjusted large gradient $\widetilde{\eta}_k^{(i)}$. If that were true, the entire proof structure of our covering LP convergence would become much closer to that of packing LP, and there would be absolutely no need for the introduction of the distance adjustment in Section 6.3, as well as the definitions of $\widetilde{A}$ and $\widetilde{\eta}$.

Unfortunately, if one replaces $\widetilde{\eta}$ with $\eta$ in the above lemma, the inequality is *false*. The reason behind it is very similar to what we have summarized in Section 6.3, and related to the fact that the exponential penalty function is not Lipschitz smooth.

### 6.7 Step 5: Putting All Together

Combining Lemma 6.12, Lemma 6.13, and Lemma 6.14, we obtain that

$$\alpha_k\big(f_\mu(\mathsf{x}_k) - f_\mu(u^*)\big) - \alpha_k\varepsilon\mathsf{OPT}$$

[13] This negative width technique is related to [8, Definition 3.2], where the authors analyze the multiplicative weight update method in a special case when the oracle returns loss values only in $[-\ell,+\rho]$, for some $\ell \ll \rho$. This technique is also a sub-case of a more general theory of mirror descent, known as the local-norm convergence, that we have summarized in a separate and later paper [4].

$$\leq \frac{(1-\tau)\alpha_k}{\tau}(f_\mu(\mathsf{y}_{k-1}) - f_\mu(\mathsf{x}_k)) + \mathbf{E}_i\Big[\alpha_k\langle n\xi_k^{(i)}, \mathsf{z}_{k-1} - u^*\rangle\Big] + \mathbf{E}_i\Big[\alpha_k\langle n\widetilde{\eta}_k^{(i)}, -u^*\rangle\Big]$$

$$\leq \frac{(1-\tau)\alpha_k}{\tau}(f_\mu(\mathsf{y}_{k-1}) - f_\mu(\mathsf{x}_k)) + V_{\mathsf{z}_{k-1}}\big(\frac{u^*}{1+\gamma}\big) - \mathbf{E}_i\Big[V_{\mathsf{z}_k^{(i)}}\big(\frac{u^*}{1+\gamma}\big)\Big]$$
$$+ 12\mathsf{OPT}\cdot\gamma\alpha_k\beta + \mathbf{E}_i\Big[\frac{12\alpha_k n}{\mu\beta}\big(f_\mu(\mathsf{x}_k) - f_\mu(\mathsf{y}_k^{(i)})\big)\Big]$$

**Remark 6.15.** Above, the quantity "$12\mathsf{OPT}\cdot\gamma\alpha_k\beta$" is the loss term introduced by the mirror descent. Unlike the packing LP case —see (3.2)— this loss term is not dominated by the gradient step. (If one could do so, this would give `CovLPSolver` an $\varepsilon^{-1}$ convergence rate.)

The quantity "$\alpha_k\langle n\xi_k^{(i)}, \mathsf{z}_{k-1}-u^*\rangle$" is the loss introduced by the (adjusted) large gradient $\widetilde{\eta}$, and is dominated by our gradient step progress owing to Lemma 6.14. This is similar to the packing LP case —see Lemma 3.10.

From here, let us use the special choice of $\tau = \frac{\mu\beta}{12n}$. We obtain that

$$-\alpha_k\big(f_\mu(u^*) + \varepsilon\mathsf{OPT}\big)$$
$$\leq 12\gamma\alpha_k\beta\mathsf{OPT} + \frac{(1-\tau)\alpha_k}{\tau}f_\mu(\mathsf{y}_{k-1}) + V_{\mathsf{z}_{k-1}}\big(\frac{u^*}{1+\gamma}\big) - \mathbf{E}_i\Big[\frac{\alpha_k}{\tau}f_\mu(\mathsf{y}_k^{(i)}) + V_{\mathsf{z}_k^{(i)}}\big(\frac{u^*}{1+\gamma}\big)\Big] \ .$$

Using the choice $\alpha_k = \frac{\alpha_{k-1}}{1-\tau}$ and telescoping the above inequality for $k = 1,\dots,T$, we have

$$-\big(\sum_{k=1}^T \alpha_k\big)\big(f_\mu(u^*) + \varepsilon\mathsf{OPT}\big) \leq \big(\sum_{k=1}^T \alpha_k\big)\cdot 12\gamma\beta\mathsf{OPT} + \frac{\alpha_0}{\tau}f_\mu(\mathsf{y}_0) + V_{\mathsf{z}_0}\big(\frac{u^*}{1+\gamma}\big) - \frac{\alpha_T}{\tau}\mathbf{E}\big[f_\mu(\mathsf{y}_T)\big] \ .$$

We compute that $\sum_{k=1}^T \alpha_k = \alpha_T \cdot \sum_{k=0}^{T-1}(1-\tau)^k = \alpha_T \cdot \frac{1-(1-\tau)^T}{\tau} < \frac{\alpha_T}{\tau}$, and recall that $\gamma = 2\alpha_T n$. Therefore, we rearrange and get

$$\frac{\alpha_T}{\tau}\mathbf{E}\big[f_\mu(\mathsf{y}_T)\big] \leq \frac{\alpha_T}{\tau}\big(f_\mu(u^*) + \varepsilon\mathsf{OPT}\big) + \frac{\alpha_T}{\tau}\cdot 12\gamma\beta\mathsf{OPT} + \frac{\alpha_0}{\tau}f_\mu(\mathsf{y}_0) + V_{\mathsf{z}_0}\big(\frac{u^*}{1+\gamma}\big) \ ,$$
$$\implies \mathbf{E}\big[f_\mu(\mathsf{y}_T)\big] \leq f_\mu(u^*) + \varepsilon\mathsf{OPT} + 24\alpha_T n\beta\mathsf{OPT} + (1-\tau)^T f_\mu(\mathsf{y}_0) + \frac{\tau}{\alpha_T}V_{\mathsf{z}_0}\big(\frac{u^*}{1+\gamma}\big) \ . \quad (6.3)$$

From this point, we need to use our special choice of the initial point $\mathsf{x}_0 = \mathsf{y}_0 = \mathsf{z}_0 = x^{\mathsf{start}}$ (see Proposition 6.2), which implies that $f_\mu(\mathsf{y}_0) \leq 4\mathsf{OPT}$ and $\mathbb{1}^T x^{\mathsf{start}} \leq 4\mathsf{OPT}$. We also have

$$V_{\mathsf{z}_0}\big(\frac{u^*}{1+\gamma}\big) = V_{x^{\mathsf{start}}}\big(\frac{u^*}{1+\gamma}\big) = \sum_{i=1}^n \frac{u_i^*}{1+\gamma}\log\frac{u_i^*}{(1+\gamma)x_i^{\mathsf{start}}} + x_i^{\mathsf{start}} - \frac{u_i^*}{1+\gamma}$$
$$\overset{①}{\leq} \sum_{i=1}^n u_i^*\log(u_i^*\cdot n) + 4\mathsf{OPT} \overset{②}{\leq} (2\log(nm)+4)\cdot\mathsf{OPT} \ .$$

Above, inequality ① follows because $x_i^{\mathsf{start}} \geq 1/n$ for all $i \in [n]$ according to the definition in Proposition 6.2; inequality ② follows because each $u_i^* \leq (1+\varepsilon/2)x_i^* \leq (1+\varepsilon/2)\mathsf{OPT} \leq (1+\varepsilon/2)m$ and $\mathbb{1}^T u_i^* = (1+\varepsilon/2)\mathsf{OPT}$, as well as the fact that $\varepsilon$ is sufficiently small.

Finally, we choose $\beta = \sqrt{\varepsilon}$, $T = \lceil\frac{1}{\tau}\log(1/\varepsilon)\rceil$, and $\alpha_0$ such that $\alpha_T = \frac{\varepsilon}{12n\beta}$. Substituting into (6.3) all of these parameters, along with the aforementioned inequalities $f_\mu(\mathsf{y}_0) \leq 4\mathsf{OPT}$ and $V_{\mathsf{z}_0}\big(\frac{u^*}{1+\gamma}\big) \leq (2\log(nm)+4)\cdot\mathsf{OPT}$, as well as $f_\mu(u^*) \leq (1+\varepsilon)\mathsf{OPT}$ from Proposition 4.5.a, we obtain that

$$\mathbf{E}\big[f_\mu(\mathsf{y}_T)\big] \leq (1+\varepsilon)\mathsf{OPT} + \varepsilon\mathsf{OPT} + 2\varepsilon\mathsf{OPT} + \varepsilon f_\mu(\mathsf{y}_0) + \frac{\mu\beta/12n}{\varepsilon/12n\beta}(2\log(nm)+4)\mathsf{OPT}$$
$$\leq (1+9\varepsilon)\mathsf{OPT} \ .$$

This finishes the proof of Theorem 6.6. □

# Acknowledgements

We would like to thank Jon Kelner, Neal Young and the anonymous reviewers for discussing this paper.

# Appendix

## A Proof of Lemma 3.6

**Lemma 3.6.** *We have* $\mathsf{x}_k, \mathsf{y}_k, \mathsf{z}_k \in \Delta_{\mathsf{box}}$ *for all* $k = 0, 1, \dots, T$.

*Proof.* This is true at the beginning as $\mathsf{x}_0 = \mathsf{y}_0 = x^{\mathsf{start}} \in \Delta_{\mathsf{box}}$ (see Fact 2.8) and $\mathsf{z}_0 = 0 \in \Delta_{\mathsf{box}}$. In fact, it suffices for us to show that for every $k \geq 1$, $\mathsf{y}_k = \sum_{l=0}^{k} \gamma_k^l \mathsf{z}_l$ for some scalars $\gamma_k^l$ satisfying $\sum_l \gamma_k^l = 1$ and $\gamma_k^l \geq 0$ for each $l = 0, \dots, k$. If this is true, we can prove the lemma by induction: at each iteration $k \geq 1$,

1. $\mathsf{x}_k = \tau \mathsf{z}_{k-1} + (1-\tau)\mathsf{y}_{k-1}$ must be in $\Delta_{\mathsf{box}}$ because $\mathsf{y}_{k-1}$ and $\mathsf{z}_{k-1}$ are and $\tau \in [0,1]$,
2. $\mathsf{z}_k$ is in $\Delta_{\mathsf{box}}$ by the definition that $\mathsf{z}_k = \arg\min_{z \in \Delta_{\mathsf{box}}}\{\cdots\}$, and
3. $\mathsf{y}_k$ is also in $\Delta_{\mathsf{box}}$ because $\mathsf{y}_k = \sum_{l=0}^{k} \gamma_k^l \mathsf{z}_l$ is a convex combination of the $\mathsf{z}_l$'s and $\Delta_{\mathsf{box}}$ is convex.

For the rest of the proof, we show that $\mathsf{y}_k = \sum_{l=0}^{k} \gamma_k^l \mathsf{z}_l$ for every $k \geq 1$ with coefficients[14]

$$\gamma_k^l = \begin{cases} (1-\tau)\gamma_{k-1}^l, & l = 0, \dots, k-2; \\ \big(\frac{1}{n\alpha_{k-1}L} - \frac{1}{n\alpha_k L}\big) + \tau\big(1 - \frac{1}{n\alpha_{k-1}L}\big), & l = k-1; \\ \frac{1}{n\alpha_k L}, & l = k. \end{cases}$$

This is true at the base case $k = 1$ because $\mathsf{y}_1 = \mathsf{x}_1 + \frac{1}{n\alpha_1 L}(\mathsf{z}_1 - \mathsf{z}_0) = \frac{1}{n\alpha_1 L}\mathsf{z}_1 + \big(1 - \frac{1}{n\alpha_1 L}\big)\mathsf{z}_0$. For the general $k \geq 2$, we have

$$\begin{aligned} \mathsf{y}_k &= \mathsf{x}_k + \frac{1}{n\alpha_k L}(\mathsf{z}_k - \mathsf{z}_{k-1}) \\ &= \tau \mathsf{z}_{k-1} + (1-\tau)\mathsf{y}_{k-1} + \frac{1}{n\alpha_k L}(\mathsf{z}_k - \mathsf{z}_{k-1}) \\ &= \tau \mathsf{z}_{k-1} + (1-\tau)\Big(\sum_{l=0}^{k-2} \gamma_{k-1}^l \mathsf{z}_l + \frac{1}{n\alpha_{k-1}L}\mathsf{z}_{k-1}\Big) + \frac{1}{n\alpha_k L}(\mathsf{z}_k - \mathsf{z}_{k-1}) \\ &= \Big(\sum_{l=0}^{k-2}(1-\tau)\gamma_{k-1}^l \mathsf{z}_l\Big) + \bigg(\Big(\frac{1}{n\alpha_{k-1}L} - \frac{1}{n\alpha_k L}\Big) + \tau\Big(1 - \frac{1}{n\alpha_{k-1}L}\Big)\bigg)\mathsf{z}_{k-1} + \frac{1}{n\alpha_k L}\mathsf{z}_k \enspace . \end{aligned}$$

Therefore, we obtain $\mathsf{y}_k = \sum_{l=0}^{k} \gamma_k^l \mathsf{z}_l$ as desired.

It is now easy to check that under our definition of $\alpha_k$ (which satisfies $\alpha_k \geq \alpha_{k-1}$ and $\alpha_k \geq \alpha_0 = \frac{1}{nL}$, we must have $\gamma_k^l \geq 0$ for all $k$ and $l$. Also,

$$\begin{aligned} \sum_l \gamma_k^l &= \sum_{l=0}^{k-2}(1-\tau)\gamma_{k-1}^l + \bigg(\Big(\frac{1}{n\alpha_{k-1}L} - \frac{1}{n\alpha_k L}\Big) + \tau\Big(1 - \frac{1}{n\alpha_{k-1}L}\Big)\bigg) + \frac{1}{n\alpha_k L} \\ &= (1-\tau)\Big(1 - \frac{1}{n\alpha_{k-1}L}\Big) + \bigg(\Big(\frac{1}{n\alpha_{k-1}L} - \frac{1}{n\alpha_k L}\Big) + \tau\Big(1 - \frac{1}{n\alpha_{k-1}L}\Big)\bigg) + \frac{1}{n\alpha_k L} = 1 \enspace . \end{aligned}$$

□

[14] We wish to point out that this proof coincides with a lemma from the accelerated coordinate descent theory of Fercoq and Richtárik [18]. Their paper is about optimizing an objective function that is Lipschitz smooth, and thus irrelevant to our work.

# B Proof of Proposition 4.5

**Proposition 4.5.**

(a) $f_\mu(u^*) \le (1+\varepsilon)\mathsf{OPT}$ *for* $u^* \stackrel{\text{def}}{=} (1+\varepsilon/2)x^*$.

(b) $f_\mu(x) \ge (1-\varepsilon)\mathsf{OPT}$ *for every* $x \ge 0$.

(c) *For any* $x \ge 0$ *satisfying* $f_\mu(x) \le 2\mathsf{OPT}$, *we must have* $Ax \ge (1-\varepsilon)\mathbb{1}$.

(d) *If* $x \ge 0$ *satisfies* $f_\mu(x) \le (1+\delta)\mathsf{OPT}$ *for some* $\delta \in [0,1]$, *then* $\frac{1}{1-\varepsilon}x$ *is a* $\frac{1+\delta}{1-\varepsilon}$*-approximate solution to the covering LP.*

*Proof.*

(a) We have $\mathbb{1}^T u^* = (1+\varepsilon/2)\mathsf{OPT}$ by the definition of $\mathsf{OPT}$. Also, from the feasibility constraint $Ax^* \ge \mathbb{1}$ in the covering LP, we have $Au^* - \mathbb{1} \ge \varepsilon/2 \cdot \mathbb{1}$, and can compute $f_\mu(u^*)$ as follows:

$$f_\mu(u^*) = \mu \sum_j e^{\frac{1}{\mu}(1-(Au^*)_j)} + \mathbb{1}^T u^* \le \mu \sum_j e^{\frac{-\varepsilon/2}{\mu}} + (1+\varepsilon/2)\mathsf{OPT}$$
$$\le \frac{\mu m}{(nm)^2} + (1+\varepsilon/2)\mathsf{OPT} \le (1+\varepsilon)\mathsf{OPT} \ .$$

(b) Suppose towards contradiction that $f_\mu(x) < (1-\varepsilon)\mathsf{OPT}$. Since $f_\mu(x) < \mathsf{OPT} \le m$, we must have that for every $j \in [m]$, it satisfies that $e^{\frac{1}{\mu}(1-(Ax)_j)} \le f_\mu(x)/\mu \le m/\mu$. This further implies $(Ax)_j \ge 1-\varepsilon$ by the definition of $\mu$. In other words, $Ax \ge (1-\varepsilon)\mathbb{1}$. By the definition of $\mathsf{OPT}$, we must then have $\mathbb{1}^T x \ge (1-\varepsilon)\mathsf{OPT}$, finishing the proof that $f_\mu(x) \ge \mathbb{1}^T x \ge (1-\varepsilon)\mathsf{OPT}$, giving a contradiction.

(c) To show $Ax \ge (1-\varepsilon)\mathbb{1}$, we can assume that $v = \max_j(1-(Ax)_j) > \varepsilon$ because otherwise we are done. Under this definition, we have

$$f_\mu(x) \ge \mu e^{\frac{v}{\mu}} = \mu\big((\tfrac{nm}{\varepsilon})^4\big)^{v/\varepsilon} \ge \tfrac{\varepsilon}{4\log(nm/\varepsilon)}(\tfrac{nm}{\varepsilon})^4 \gg 2\mathsf{OPT} \ ,$$

contradicting to our assumption that $f_\mu(x) \le 2\mathsf{OPT}$. Therefore, we must have $v \le \varepsilon$, that is, $Ax \ge (1-\varepsilon)\mathbb{1}$.

(d) For any $x$ satisfying $f_\mu(x) \le (1+\theta)\mathsf{OPT} \le 2\mathsf{OPT}$, owing to Proposition 4.5.c, we first have that $x$ is approximately feasible, i.e., $Ax \ge (1-\varepsilon)\mathbb{1}$. Next, because $\mathbb{1}^T x \le f_\mu(x) \le (1+\theta)\mathsf{OPT}$, we know that $x$ yields an objective $\mathbb{1}^T x \le (1+\theta)\mathsf{OPT}$. Letting $x' = \frac{1}{1-\varepsilon}x$, we both have that $x'$ is feasible (i.e., $Ax' \ge \mathbb{1}$), and $x'$ has an objective $\mathbb{1}^T x'$ at most $\frac{1+\delta}{1-\varepsilon}\mathsf{OPT}$. □

# C Missing Proofs for Section 5

In this section we prove Theorem 5.3. Because the proof structure is almost identical to that of Theorem 3.4, we spend most of the discussions only pointing out the difference rather than repeating the proofs. The following three lemmas are completely identical to the ones in the packing LP case, so we restate them below:

**Lemma C.1** (cf. Lemma 3.3)**.** *Each iteration of* `CovLPSolver`$^{\mathsf{wb}}$ *can be implemented to run in expected* $O(N/n)$ *time.*

**Lemma C.2** (cf. Lemma 3.6)**.** *We have* $\mathsf{x}_k, \mathsf{y}_k, \mathsf{z}_k \in \Delta_{\mathsf{box}}$ *for all* $k = 0, 1, \dots, T$.

**Lemma C.3** (cf. Lemma 3.7)**.** *For every* $u \in \Delta_{\mathsf{box}}$, *it satisfies* $\langle n\alpha_k \xi_k^{(i)}, \mathsf{z}_{k-1} - u\rangle \le n^2\alpha_k^2 L \cdot \langle \xi_k^{(i)}, \mathsf{x}_k - \mathsf{y}_k^{(i)}\rangle + \frac{1}{2}\|\mathsf{z}_{k-1} - u\|_A^2 - \frac{1}{2}\|\mathsf{z}_k^{(i)} - u\|_A^2$ .

For the gradient descent guarantee of Section 3.3, one can first note that Lemma 2.7 remains true: this can be verified by replacing $\nabla_i f_\mu(x) + 1$ in its proof with $1 - \nabla_i f_\mu(x)$. For this reason, Lemma 3.9 (which is built on Lemma 2.7) also remains true. We state it below:

**Lemma C.4** (cf. Lemma 3.9)**.** *We have* $f_\mu(\mathsf{x}_k) - f_\mu(\mathsf{y}_k^{(i)}) \geq \frac{1}{2}\langle \nabla f_\mu(\mathsf{x}_k), \mathsf{x}_k - \mathsf{y}_k^{(i)} \rangle \geq 0$.

**Putting All Together.** Denote by $\eta_k^{(i)} \in \mathbb{R}^n_{\leq 0}$ the vector that is only non-zero at coordinate $i$, and satisfies $\eta_{k,i}^{(i)} = \nabla_i f_\mu(\mathsf{x}_k) - \xi_{k,i}^{(i)} \in (-\infty, 0]$. In other words, the full gradient

$$\nabla f_\mu(\mathsf{x}_k) = \mathbf{E}_i[(0, \ldots, n\nabla_i f_\mu(\mathsf{x}_k), \ldots, 0)] = \mathbf{E}_i[n\eta_k^{(i)} + n\xi_k^{(i)}]$$

can be (in expectation) decomposed into the a large but non-positive component $\eta_k^{(i)} \in (-\infty, 0]^n$ and a small component $\xi_k^{(i)} \in [-1, 1]^n$. Similar as Section 3.4, for any $u \in \Delta_{\mathsf{box}}$, we can use a basic convexity argument and the mirror descent lemma to compute that

$$\begin{aligned}
&\alpha_k(f_\mu(\mathsf{x}_k) - f_\mu(u)) \leq \langle \alpha_k \nabla f_\mu(\mathsf{x}_k), \mathsf{x}_k - u \rangle \\
&= \langle \alpha_k \nabla f_\mu(\mathsf{x}_k), \mathsf{x}_k - \mathsf{z}_{k-1} \rangle + \langle \alpha_k \nabla f_\mu(\mathsf{x}_k), \mathsf{z}_{k-1} - u \rangle \\
&= \langle \alpha_k \nabla f_\mu(\mathsf{x}_k), \mathsf{x}_k - \mathsf{z}_{k-1} \rangle + \mathbf{E}_i\Big[\langle n\alpha_k \eta_k^{(i)}, \mathsf{z}_{k-1} - u \rangle + \langle n\alpha_k \xi_k^{(i)}, \mathsf{z}_{k-1} - u \rangle\Big] \\
&\overset{①}{=} \frac{(1-\tau)\alpha_k}{\tau}\langle \nabla f_\mu(\mathsf{x}_k), \mathsf{y}_{k-1} - \mathsf{x}_k \rangle + \mathbf{E}_i\Big[\langle n\alpha_k \eta_k^{(i)}, \mathsf{z}_{k-1} - u \rangle + \langle n\alpha_k \xi_k^{(i)}, \mathsf{z}_{k-1} - u \rangle\Big] \qquad (C.1) \\
&\overset{②}{\leq} \frac{(1-\tau)\alpha_k}{\tau}(f_\mu(\mathsf{y}_{k-1}) - f_\mu(\mathsf{x}_k)) \\
&\quad + \mathbf{E}_i\Big[\boxed{\langle n\alpha_k \eta_k^{(i)}, \mathsf{z}_{k-1} - u \rangle + n^2\alpha_k^2 L \cdot \langle \xi_k^{(i)}, \mathsf{x}_k - \mathsf{y}_k^{(i)} \rangle} + \frac{1}{2}\|\mathsf{z}_{k-1} - u\|_A^2 - \frac{1}{2}\|\mathsf{z}_k^{(i)} - u\|_A^2\Big] \qquad (C.2)
\end{aligned}$$

Above, ① is because $\mathsf{x}_k = \tau \mathsf{z}_{k-1} + (1-\tau)\mathsf{y}_{k-1}$, which implies that $\tau(\mathsf{x}_k - \mathsf{z}_{k-1}) = (1-\tau)(\mathsf{y}_{k-1} - \mathsf{x}_k)$. ② uses convexity and Lemma C.3. We can establish the following lemma to upper bound the boxed term in (C.2). Its proof is in the same spirit to that of Lemma 3.10, and is the *only* place that we require all vectors to reside in $\Delta_{\mathsf{box}}$.

**Lemma C.5** (cf. Lemma 3.10)**.** *For every* $u \in \Delta_{\mathsf{box}}$,

$$\langle n\alpha_k \eta_k^{(i)}, \mathsf{z}_{k-1} - u \rangle + n^2\alpha_k^2 L \cdot \langle \xi_k^{(i)}, \mathsf{x}_k - \mathsf{y}_k^{(i)} \rangle \leq 21 n\alpha_k L \cdot (f_\mu(\mathsf{x}_k) - f_\mu(\mathsf{y}_k^{(i)})) \ .$$

*Proof of Lemma C.5.* Now there are three possibilities:

- If $\eta_{k,i}^{(i)} = 0$, then we must have $\xi_{k,i}^{(i)} = \nabla_i f_\mu(\mathsf{x}_k) \in [-1, 1]$. Lemma C.4 implies

$$\langle n\alpha_k \eta_k^{(i)}, \mathsf{z}_{k-1} - u \rangle + n^2\alpha_k^2 L \cdot \langle \xi_k^{(i)}, \mathsf{x}_k - \mathsf{y}_k^{(i)} \rangle = n^2\alpha_k^2 L \cdot \langle \nabla f_\mu(\mathsf{x}_k), \mathsf{x}_k - \mathsf{y}_k^{(i)} \rangle \leq 2n^2\alpha_k^2 L \cdot (f_\mu(\mathsf{x}_k) - f_\mu(\mathsf{y}_k^{(i)}))$$

- If $\eta_{k,i}^{(i)} < 0$ and $\mathsf{z}_{k,i}^{(i)} < \frac{10}{\|A_{:i}\|_\infty}$ (thus $\mathsf{z}_k^{(i)}$ is not on the boundary of $\Delta_{\mathsf{box}}$), then we precisely have $\mathsf{z}_{k,i}^{(i)} = \mathsf{z}_{k-1,i} + \frac{n\alpha_k}{\|A_{:i}\|_\infty}$, and accordingly $\mathsf{y}_{k,i}^{(i)} = \mathsf{x}_{k,i} + \frac{1}{L\|A_{:i}\|_\infty} > \mathsf{x}_{k,i}$. In this case,

$$\begin{aligned}
&\langle n\alpha_k \eta_k^{(i)}, \mathsf{z}_{k-1} - u \rangle + n^2\alpha_k^2 L \cdot \langle \xi_k^{(i)}, \mathsf{x}_k - \mathsf{y}_k^{(i)} \rangle \\
&\overset{①}{\leq} n\alpha_k \cdot \nabla_i f_\mu(\mathsf{x}_k) \cdot \frac{-10}{\|A_{:i}\|_\infty} + n^2\alpha_k^2 L \cdot \langle \xi_k^{(i)}, \mathsf{x}_k - \mathsf{y}_k^{(i)} \rangle \\
&\overset{②}{<} n\alpha_k \cdot \nabla_i f_\mu(\mathsf{x}_k) \cdot \frac{-10}{\|A_{:i}\|_\infty} + n^2\alpha_k^2 L \cdot \langle \nabla f_\mu(\mathsf{x}_k), \mathsf{x}_k - \mathsf{y}_k^{(i)} \rangle \\
&\overset{③}{=} 10 n\alpha_k L \cdot \langle \nabla f_\mu(\mathsf{x}_k), \mathsf{x}_k - \mathsf{y}_k^{(i)} \rangle + n^2\alpha_k^2 L \cdot \langle \nabla f_\mu(\mathsf{x}_k), \mathsf{x}_k - \mathsf{y}_k^{(i)} \rangle \\
&\overset{④}{\leq} \big(20 n\alpha_k L + 2n^2\alpha_k^2 L\big) \cdot (f_\mu(\mathsf{x}_k) - f_\mu(\mathsf{y}_k^{(i)})) \ .
\end{aligned}$$

Above, ① follows from the fact that $\mathsf{z}_{k-1}, u \in \Delta_{\mathsf{box}}$ and therefore $z_{k-1,i} \geq 0$ and $u_i \leq \frac{10}{\|A_{:i}\|_\infty}$ by the definition of $\Delta_{\mathsf{box}}$, and $u \geq 0$; ② follows from the fact that $\mathsf{x}_k$ and $\mathsf{y}_k^{(i)}$ are only different at coordinate $i$, and $\xi_{k,i}^{(i)} = -1 > \nabla_i f_\mu(\mathsf{x}_k)$ (since $\eta_{k,i}^{(i)} < 0$); ③ follows from the fact that $\mathsf{y}_k^{(i)} = \mathsf{x}_k + \frac{\mathbf{e}_i}{L\|A_{:i}\|_\infty}$; and ④ uses Lemma C.4.

- If $\eta_{k,i}^{(i)} < 0$ and $\mathsf{z}_{k,i}^{(i)} = \frac{10}{\|A_{:i}\|_\infty}$, then we have

$$\begin{aligned}&\langle n\alpha_k \eta_k^{(i)}, \mathsf{z}_{k-1} - u\rangle + n^2\alpha_k^2 L \cdot \langle \xi_k^{(i)}, \mathsf{x}_k - \mathsf{y}_k^{(i)}\rangle \\ &\overset{①}{\leq} \langle n\alpha_k \eta_k^{(i)}, \mathsf{z}_{k-1} - \mathsf{z}_k^{(i)}\rangle + n^2\alpha_k^2 L \cdot \langle \nabla f_\mu(\mathsf{x}_k), \mathsf{x}_k - \mathsf{y}_k^{(i)}\rangle \\ &\overset{②}{\leq} \langle n\alpha_k \nabla f_\mu(\mathsf{x}_k), \mathsf{z}_{k-1} - \mathsf{z}_k^{(i)}\rangle + n^2\alpha_k^2 L \cdot \langle \nabla f_\mu(\mathsf{x}_k), \mathsf{x}_k - \mathsf{y}_k^{(i)}\rangle \\ &\overset{③}{=} n^2\alpha_k^2 L \cdot \langle \nabla f_\mu(\mathsf{x}_k), \mathsf{x}_k - \mathsf{y}_k^{(i)}\rangle + n^2\alpha_k^2 L \cdot \langle \nabla f_\mu(\mathsf{x}_k), \mathsf{x}_k - \mathsf{y}_k^{(i)}\rangle \overset{④}{\leq} 4n^2\alpha_k^2 L \cdot (f_\mu(\mathsf{x}_k) - f_\mu(\mathsf{y}_k^{(i)})) \ .\end{aligned}$$

  Above, ① is because $u_i \leq \frac{10}{\|A_{:i}\|_\infty} = \mathsf{z}_{k,i}^{(i)}$ and $\eta_{k,i}^{(i)} < 0$, together with $\nabla_i f_\mu(\mathsf{x}_k) < \xi_{k,i}^{(i)}$ and $\mathsf{x}_{k,i} \leq \mathsf{y}_{k,i}^{(i)}$; ② uses $\nabla_i f_\mu(\mathsf{x}_k) = \eta_{k,i}^{(i)} - 1 < \eta_{k,i}^{(i)}$ and $\mathsf{z}_{k,i}^{(i)} \geq \mathsf{z}_{k-1,i}$; ③ is from our choice of $\mathsf{y}_k$ which satisfies that $\mathsf{z}_{k-1} - \mathsf{z}_k^{(i)} = n\alpha_k L(\mathsf{x}_k - \mathsf{y}_k^{(i)})$; and ④ uses Lemma C.4.

Combining the three cases, and using the fact that $f_\mu(\mathsf{x}_k) - f_\mu(\mathsf{y}_k^{(i)}) \geq 0$, we conclude that

$$\begin{aligned}\langle n\alpha_k \eta_k^{(i)}, \mathsf{z}_{k-1} - u\rangle + n^2\alpha_k^2 L \cdot \langle \xi_k^{(i)}, \mathsf{x}_k - \mathsf{y}_k^{(i)}\rangle &\leq (20n\alpha_k L + 4n^2\alpha_k^2 L) \cdot (f_\mu(\mathsf{x}_k) - f_\mu(\mathsf{y}_k^{(i)})) \\ &\leq 21 n\alpha_k L \cdot (f_\mu(\mathsf{x}_k) - f_\mu(\mathsf{y}_k^{(i)})) \ .\end{aligned}$$

Above, the last inequality uses our choice of $\alpha_k$, which implies $n\alpha_k \leq n\alpha_T = \frac{1}{\varepsilon L} \leq \frac{1}{4}$. □

Plugging Lemma C.5 back to (C.2), we have

$$\begin{aligned}&\alpha_k(f_\mu(\mathsf{x}_k) - f_\mu(u)) \leq \langle \alpha_k \nabla f_\mu(\mathsf{x}_k), \mathsf{x}_k - u\rangle \\ &\overset{①}{\leq} \frac{(1-\tau)\alpha_k}{\tau}(f_\mu(\mathsf{y}_{k-1}) - f_\mu(\mathsf{x}_k)) + \mathbf{E}_i\Big[21n\alpha_k L \cdot (f_\mu(\mathsf{x}_k) - f_\mu(\mathsf{y}_k^{(i)})) + \frac{1}{2}\|\mathsf{z}_{k-1} - u\|_A^2 - \frac{1}{2}\|\mathsf{z}_k - u\|_A^2\Big] \\ &\overset{②}{\leq} \alpha_k f_\mu(\mathsf{x}_k) + \big(21n\alpha_k L - \alpha_k\big) f_\mu(\mathsf{y}_{k-1}) + \mathbf{E}_i\Big[-21n\alpha_k L \cdot f_\mu(\mathsf{y}_k^{(i)}) + \frac{1}{2}\|\mathsf{z}_{k-1} - u\|_A^2 - \frac{1}{2}\|\mathsf{z}_k - u\|_A^2\Big] \ .\end{aligned} \tag{C.3}$$

Above, ① uses Lemma C.5; and ② is because we have chosen $\tau$ to satisfy $\frac{1}{\tau} = 21nL$.

Next, recall that we have picked $\alpha_k$ so that $(21nL - 1)\alpha_k = 21nL \cdot \alpha_{k-1}$ in $\texttt{CovLPSolver}^{\mathsf{wb}}$. Telescoping (C.3) for $k = 1, \ldots, T$ and choosing $u^* = (1 + \varepsilon/2)x^*$, we have

$$-\sum_{k=1}^{T} \alpha_k f_\mu(u^*) \leq 21 f_\mu(\mathsf{y}_0) - 21n\alpha_T L \cdot \mathbf{E}[f_\mu(\mathsf{y}_T)] + \|\mathsf{z}_0 - u^*\|_A^2 \leq -21n\alpha_T L \cdot \mathbf{E}[f_\mu(\mathsf{y}_T)] + 75\mathsf{OPT} \ .$$

Here, the second inequality is due to $f_\mu(\mathsf{y}_0) = f_\mu(x^{\mathsf{start}}) \leq 3\mathsf{OPT}$ from Fact 5.2, and the fact that

$$\|\mathsf{z}_0 - u^*\|_A^2 = \|u^*\|_A^2 = \sum_{i=1}^{n} (u_i^*)^2 \cdot \|A_{:i}\|_\infty \leq (1+\varepsilon/2)^2 \sum_{i=1}^{n} (x_i^*)^2 \cdot \|A_{:i}\|_\infty \leq 10(1+\varepsilon/2)^2 \sum_{i=1}^{n} x_i^* < 12\mathsf{OPT} \ .$$

Finally, using the fact that $\sum_{k=1}^{T} \alpha_k = \alpha_T \cdot \sum_{k=0}^{T-1} \big(1 - \frac{1}{21nL}\big)^k = 21n\alpha_T L\big(1 - (1 - \frac{1}{21nL})^T\big)$, we rearrange and obtain that

$$\mathbf{E}[f_\mu(\mathsf{y}_T)] \leq \frac{\sum_k \alpha_k}{21n\alpha_T L} f_\mu(u^*) + \frac{75}{21n\alpha_T L}\mathsf{OPT} = \big(1 - (1 - \frac{1}{21nL})^T\big) f_\mu(u^*) + \frac{75}{21n\alpha_T L}\mathsf{OPT} \ .$$

We choose $T = \lceil 21nL\log(1/\varepsilon)\rceil$ so that $\frac{1}{n\alpha_T L} = (1-\frac{1}{21nL})^T \leq \varepsilon$. Combining this with the fact that $f_\mu(u^*) \leq (1+\varepsilon)\mathsf{OPT}$ (see Proposition 4.5.a), we obtain

$$\mathbf{E}[f_\mu(\mathsf{y}_T)] \leq (1+\varepsilon)\mathsf{OPT} + 3.6\varepsilon\cdot\mathsf{OPT} < (1+4.6\varepsilon)\mathsf{OPT} \ .$$

Therefore, we have finished proving Theorem 5.3. □

# D Missing Proofs for Section 6

**Proposition 6.4.** *If $\mathsf{z}_{k-1} \in \Delta_{\mathsf{simplex}}$ and $\mathsf{z}_{k-1} > 0$, the minimizer $z = \arg\min_{z\in\Delta_{\mathsf{simplex}}} \big\{V_{\mathsf{z}_{k-1}}(z) + \langle \delta \mathbf{e}_i, z\rangle\big\}$ for any scalar $\delta\in\mathbb{R}$ and basis vector $\mathbf{e}_i$ can be computed as follows:*

1. *$z \leftarrow \mathsf{z}_{k-1}$.*
2. *$z_i \leftarrow z_i \cdot e^{-\delta}$.*
3. *If $\mathbb{1}^T z > 2\mathsf{OPT}'$, $z \leftarrow \frac{2\mathsf{OPT}'}{\mathbb{1}^T z} z$.*
4. *Return $z$.*

*Proof.* Let us denote by $z$ the returned value of the described procedure, and $g(u) \stackrel{\mathrm{def}}{=} V_{\mathsf{z}_{k-1}}(u) + \langle \delta\mathbf{e}_i, u\rangle$. Since $\Delta_{\mathsf{simplex}}$ is a convex body and $g(\cdot)$ is convex, to show $z = \arg\min_{z\in\Delta_{\mathsf{simplex}}}\{g(u)\}$, it suffices for us to prove that for every $u\in\Delta_{\mathsf{simplex}}$, $\langle\nabla g(z), u - z\rangle \geq 0$. Since the gradient $\nabla g(z)$ can be written explicitly, this is equivalent to

$$\delta(u_i - z_i) + \textstyle\sum_{\ell=1}^n \log\frac{z_\ell}{\mathsf{z}_{k-1,\ell}}\cdot(u_\ell - z_\ell) \geq 0 \ .$$

If the re-scaling in step 3 is not executed, then we have $z_\ell = \mathsf{z}_{k-1,\ell}$ for every $\ell\neq i$, and $z_i = \mathsf{z}_{k-1,i}\cdot e^{-\delta}$; thus, the left-hand side is zero so the above inequality is true for every $u\in\Delta_{\mathsf{simplex}}$.

Otherwise, we have $\mathbb{1}^T z = 2\mathsf{OPT}'$ and there exists some constant $Z>1$ such that, $z_\ell = \mathsf{z}_{k-1,\ell}/Z$ for every $\ell\neq i$, and $z_i = \mathsf{z}_{k-1,i}\cdot e^{-\delta}/Z$. In such a case, the left-hand side equals to

$$(u_i - z_i)\cdot(\delta-\delta) + \textstyle\sum_{\ell=1}^n -\log Z\cdot(u_\ell - z_\ell) \ .$$

It is clear at this moment that since $\log Z > 0$ and $\mathbb{1}^T u \leq 2\mathsf{OPT}' = \mathbb{1}^T z$, the above quantity is always non-negative, finishing the proof. □

**Lemma 6.13.** *Denoting by $\gamma \stackrel{\mathrm{def}}{=} 2\alpha_T n$, we have*

$$\mathbf{E}_i\big[\alpha_k\langle n\xi_k^{(i)}, \mathsf{z}_{k-1} - u^*\rangle\big] \leq V_{\mathsf{z}_{k-1}}\Big(\frac{u^*}{1+\gamma}\Big) - \mathbf{E}_i\Big[V_{\mathsf{z}_k^{(i)}}\Big(\frac{u^*}{1+\gamma}\Big)\Big] + 12\mathsf{OPT}\cdot\gamma\alpha_k\beta \ .$$

*Proof.* Define $w(x) \stackrel{\mathrm{def}}{=} \sum_i x_i\log(x_i) - x_i$ and accordingly, $V_x(y) = w(y) - \langle\nabla w(x), y - x\rangle - w(x) = \sum_i y_i\log\frac{y_i}{x_i} + x_i - y_i$. We first compute using the classical analysis of mirror descent step as follows:

$$\begin{aligned}
&\gamma\alpha_k\langle n\xi_k^{(i)}, \mathsf{z}_{k-1}\rangle + \alpha_k\langle n\xi_k^{(i)}, \mathsf{z}_{k-1} - u^*\rangle \\
&= (1+\gamma)\alpha_k\Big\langle n\xi_k^{(i)}, \mathsf{z}_k^{(i)} - \frac{u^*}{1+\gamma}\Big\rangle + (1+\gamma)\alpha_k\langle n\xi_k^{(i)}, \mathsf{z}_{k-1} - \mathsf{z}_k^{(i)}\rangle \\
&\stackrel{①}{\leq} \Big\langle \nabla w(\mathsf{z}_{k-1}) - \nabla w(\mathsf{z}_k^{(i)}), \mathsf{z}_k^{(i)} - \frac{u^*}{1+\gamma}\Big\rangle + (1+\gamma)\alpha_k\langle n\xi_k^{(i)}, \mathsf{z}_{k-1} - \mathsf{z}_k^{(i)}\rangle \\
&= \Big(w\Big(\frac{u^*}{1+\gamma}\Big) - w(\mathsf{z}_{k-1}) - \Big\langle\nabla w(\mathsf{z}_{k-1}), \frac{u^*}{1+\gamma} - \mathsf{z}_{k-1}\Big\rangle\Big) \\
&\quad - \Big(w\Big(\frac{u^*}{1+\gamma}\Big) - w(\mathsf{z}_k^{(i)}) - \Big\langle\nabla w(\mathsf{z}_k^{(i)}), \frac{u^*}{1+\gamma} - \mathsf{z}_k^{(i)}\Big\rangle\Big) \\
&\quad + \Big(w(\mathsf{z}_{k-1}) - w(\mathsf{z}_k^{(i)}) - \langle\nabla w(\mathsf{z}_{k-1}), \mathsf{z}_{k-1} - \mathsf{z}_k^{(i)}\rangle\Big) + (1+\gamma)\alpha_k\langle n\xi_k^{(i)}, \mathsf{z}_{k-1} - \mathsf{z}_k^{(i)}\rangle \\
&= V_{\mathsf{z}_{k-1}}\Big(\frac{u^*}{1+\gamma}\Big) - V_{\mathsf{z}_k^{(i)}}\Big(\frac{u^*}{1+\gamma}\Big) + \boxed{(1+\gamma)\alpha_k\langle n\xi_k^{(i)}, \mathsf{z}_{k-1} - \mathsf{z}_k^{(i)}\rangle - V_{\mathsf{z}_{k-1}}(\mathsf{z}_k^{(i)})} \ .
\end{aligned} \tag{D.1}$$

Above, ① is because $\mathsf{z}_k^{(i)} = \arg\min_{z\in\Delta_{\mathsf{simplex}}} \{V_{\mathsf{z}_{k-1}}(z) + \langle(1+\gamma)\alpha_k n\xi_k^{(i)}, z\rangle\}$, which is equivalent to saying

$$\begin{aligned} &\forall u \in \Delta_{\mathsf{simplex}}, \quad \langle \nabla V_{\mathsf{z}_{k-1}}(\mathsf{z}_k^{(i)}) + (1+\gamma)\alpha_k n\xi_k^{(i)}, u - \mathsf{z}_k^{(i)}\rangle \geq 0 \\ \iff\ &\forall u \in \Delta_{\mathsf{simplex}}, \quad \langle \nabla w(\mathsf{z}_k^{(i)}) - \nabla w(\mathsf{z}_{k-1}) + (1+\gamma)\alpha_k n\xi_k^{(i)}, u - \mathsf{z}_k^{(i)}\rangle \geq 0 \ . \end{aligned}$$

In particular, we have $\mathbb{1}^T \frac{u^*}{1+\gamma} = \mathbb{1}^T \frac{(1+\varepsilon/2)x^*}{1+\gamma} < 2\mathsf{OPT} \leq 2\mathsf{OPT}'$ and therefore $\frac{u^*}{1+\gamma} \in \Delta_{\mathsf{simplex}}$. Substituting $u = \frac{u^*}{1+\gamma}$ into the above inequality we get ①.

Next, we upper bound the term in the box:

$$\begin{aligned} &(1+\gamma)\alpha_k\langle n\xi_k^{(i)}, \mathsf{z}_{k-1} - \mathsf{z}_k^{(i)}\rangle - V_{\mathsf{z}_{k-1}}(\mathsf{z}_k^{(i)}) \\ &\overset{①}{\leq} (1+\gamma)\alpha_k n\xi_{k,i}\cdot(\mathsf{z}_{k-1,i} - \mathsf{z}_{k,i}^{(i)}) - \Big(\mathsf{z}_{k,i}^{(i)} \log\frac{\mathsf{z}_{k,i}^{(i)}}{\mathsf{z}_{k-1,i}} + \mathsf{z}_{k-1,i} - \mathsf{z}_{k,i}^{(i)}\Big) \\ &\overset{②}{\leq} (1+\gamma)\alpha_k n\xi_{k,i}\cdot(\mathsf{z}_{k-1,i} - \mathsf{z}_{k,i}^{(i)}) - \frac{|\mathsf{z}_{k,i}^{(i)} - \mathsf{z}_{k-1,i}|^2}{2\max\{\mathsf{z}_{k,i}^{(i)}, \mathsf{z}_{k-1,i}\}} \\ &\overset{③}{\leq} (1+\gamma)\alpha_k n\xi_{k,i}\cdot(\mathsf{z}_{k-1,i} - \mathsf{z}_{k,i}^{(i)}) - \frac{|\mathsf{z}_{k,i}^{(i)} - \mathsf{z}_{k-1,i}|^2}{4\mathsf{z}_{k-1,i}} \\ &\overset{④}{\leq} (1+\gamma)^2\mathsf{z}_{k-1,i}\cdot(\alpha_k n\xi_{k,i})^2 \overset{⑤}{\leq} 2\mathsf{z}_{k-1,i}\cdot(\alpha_k n\xi_{k,i})^2 \overset{⑥}{\leq} \mathsf{z}_{k-1,i}\cdot\gamma\alpha_k n|\xi_{k,i}| \\ &\overset{⑦}{\leq} \mathsf{z}_{k-1,i}\cdot\gamma\alpha_k n\xi_{k,i} + 2\mathsf{z}_{k-1,i}\cdot\gamma\alpha_k n\beta = \gamma\alpha_k\langle n\xi_k^{(i)}, \mathsf{z}_{k-1}\rangle + 2\mathsf{z}_{k-1,i}\cdot\gamma\alpha_k n\beta \ . \end{aligned} \tag{D.2}$$

Above, ① uses the facts (i) $a\log\frac{a}{b} + b - a \geq 0$ for any $a, b > 0$, (ii) $\mathsf{z}_{k-1,i} - \mathsf{z}_k^{(i)}$ and $\xi_{k,i}$ have the same sign, and (iii) $\xi_{k,i'}^{(i)} = 0$ for every $i' \neq i$; ② uses the inequality that for every $a, b > 0$, we have $a\log\frac{a}{b} + b - a \geq \frac{(a-b)^2}{2\max\{a,b\}}$.[15] ③ uses the fact that $\mathsf{z}_{k,i}^{(i)} \leq 2\mathsf{z}_{k-1,i}$.[16] ④ uses Cauchy-Shwarz: $ab - b^2/4 \leq a^2$. ⑤ uses $(1+\gamma)^2 < 2$. ⑥ uses $|\xi_{k,i}| \leq 1$ and $\gamma = 2\alpha_T n \geq 2\alpha_k n$. ⑦ uses $\xi_{k,i} \geq -\beta$.

Next, we combine (D.1) and (D.2) to conclude that

$$\alpha_k\big\langle n\xi_k^{(i)}, \mathsf{z}_{k-1} - u^*\big\rangle \leq V_{\mathsf{z}_{k-1}}\big(\frac{u^*}{1+\gamma}\big) - V_{\mathsf{z}_k^{(i)}}\big(\frac{u^*}{1+\gamma}\big) + 2\mathsf{z}_{k-1,i}\cdot\gamma\alpha_k n\beta \ .$$

Taking expectation on both sides with respect to $i$, and using the property that $\mathbb{1}^T\mathsf{z}_{k-1} \leq 3\mathsf{OPT}' \leq 6\mathsf{OPT}$, we obtain that

$$\mathbf{E}_i\big[\alpha_k\big\langle n\xi_k^{(i)}, \mathsf{z}_{k-1} - u^*\big\rangle\big] \leq V_{\mathsf{z}_{k-1}}\big(\frac{u^*}{1+\gamma}\big) - \mathbf{E}_i\Big[V_{\mathsf{z}_k^{(i)}}\big(\frac{u^*}{1+\gamma}\big)\Big] + 12\mathsf{OPT}\cdot\gamma\alpha_k\beta \ . \qquad \square$$

**Lemma 6.14.** *For every $i \in [n]$, we have*

*(a)* $f_\mu(\mathsf{x}_k) - f_\mu(\mathsf{y}_k^{(i)}) \geq 0$, *and*

*(b)* $f_\mu(\mathsf{x}_k) - f_\mu(\mathsf{y}_k^{(i)}) \geq \frac{\mu\beta}{12}\cdot\langle -\widetilde{\eta}_k^{(i)}, u^*\rangle$ .

*Proof of Lemma 6.14 part (a).* Since if $i \notin B_k$ is not a large index we have $\mathsf{y}_k^{(i)} = \mathsf{x}_k$ and the claim is trivial, we focus on $i \in B_k$ in the remaining proof. Recall that $\mathsf{y}_k^{(i)} = \mathsf{x}_k + \delta\mathbf{e}_i$ for some $\delta > 0$ defined in Algorithm 3, so we have

$$f_\mu(\mathsf{x}_k) - f_\mu(\mathsf{y}_k^{(i)}) = \int_{\tau=0}^{\delta}\langle -\nabla f_\mu(\mathsf{x}_k + \tau\mathbf{e}_i), \mathbf{e}_i\rangle d\tau = \int_{\tau=0}^{\delta}\big(\langle A_{:i}, p(\mathsf{x}_k + \tau\mathbf{e}_i)\rangle - 1\big)d\tau \ .$$

[15] This inequality in fact corresponds to a local strong convexity property of $w(\cdot)$. We have used this technique in our paper [2].

[16] This is because, our parameter choices ensure that $(1+\gamma)\alpha_k n < 1/2\beta$, which further means $-(1+\gamma)\alpha_k n\xi_{k,i}^{(i)} \leq 1/2$. As a result, we must have $\mathsf{z}_{k,i}^{(i)} \leq \mathsf{z}_{k-1,i}\cdot e^{0.5} < 2\mathsf{z}_{k-1,i}$ (see the explicit definition of the mirror step at Proposition 6.4).

It is clear that $\langle A_{:i}, p(\mathsf{x}_k + \tau \mathbf{e}_i)\rangle$ decreases as $\tau$ increases, and therefore it suffices to prove that $\langle A_{:i}, p(\mathsf{x}_k + \delta \mathbf{e}_i)\rangle \geq 1$.

Suppose that the rows of $A_{:i}$ are sorted (for the simplicity of notation) by the increasing order of $A_{j,i}$. Now, by the definition of the algorithm (recall (6.1)), there exists some $j^* \in [m]$ satisfying that

$$\sum_{j<j^*} A_{j,i} \cdot p_j(\mathsf{x}_k) < 1+\beta \quad \text{and} \quad \sum_{j\leq j^*} A_{j,i} \cdot p_j(\mathsf{x}_k) \geq 1+\beta \ .$$

Next, by our choice of $\delta$ which satisfies $\delta = \frac{\mu\beta}{2A_{j^*,i}} \leq \frac{\mu\beta}{2A_{j,i}}$ for every $j \leq j^*$, we have for every $j \leq j^*$:

$$p_j(\mathsf{x}_k + \delta \mathbf{e}_i) = p_j(\mathsf{x}_k) \cdot e^{-\frac{A_{j,i}\delta}{\mu}} \geq p_j(\mathsf{x}_k) \cdot e^{-\beta/2} \geq p_j(\mathsf{x}_k) \cdot (1-\beta/2) \ ,$$

and as a result,

$$\langle A_{:i}, p(\mathsf{x}_k + \delta \mathbf{e}_i) \geq \sum_{j\leq j^*} A_{j,i} \cdot p_j(\mathsf{x}_k + \delta \mathbf{e}_i) \geq (1-\beta/2) \sum_{j\leq j^*} A_{j,i} \cdot p_j(\mathsf{x}_k) \geq (1-\beta/2)(1+\beta) \geq 1 \ . \ \square$$

*Proof of Lemma 6.14 part (b).* Owing to part (a), for every coordinate $i$ such that $\widetilde{\eta}_{k,i} \geq 0$, we automatically have $f_\mu(\mathsf{x}_k) - f_\mu(\mathsf{y}_k^{(i)}) \geq 0$ so the lemma is obvious. Therefore, let us focus only on coordinates $i$ such that $\widetilde{\eta}_{k,i} < 0$; these are necessarily large indices $i \in B$. Recall from Definition 6.11 that $\widetilde{\eta}_{k,i} = (1+\beta) - (\widetilde{A}^T p(\mathsf{x}_k))_i$, so we have

$$\textstyle\sum_{j=1}^m \widetilde{A}_{j,i} \cdot p_j(\mathsf{x}_k) - (1+\beta) > 0 \ .$$

For the simplicity of description, suppose again that each $i$-th column is sorted in non-decreasing order, that is, $A_{1,i} \leq \cdots \leq A_{m,i}$. The definition of $j^*$ can be simplified as

$$\textstyle\sum_{j<j^*} A_{j,i} \cdot p_j(\mathsf{x}_k) < 1+\beta \quad \text{and} \quad \sum_{j\leq j^*} A_{j,i} \cdot p_j(\mathsf{x}_k) \geq 1+\beta \ .$$

Let $j^\flat \in [m]$ be the row such that

$$\textstyle\sum_{j<j^\flat} \widetilde{A}_{j,i} \cdot p_j(\mathsf{x}_k) < 1+\beta \quad \text{and} \quad \sum_{j\leq j^\flat} \widetilde{A}_{j,i} \cdot p_j(\mathsf{x}_k) \geq 1+\beta \ .$$

Note that such a $j^\flat$ must exist because $\sum_{j=1}^m \widetilde{A}_{j,i} \cdot p_j > 1+\beta$. It is clear that $j^\flat \geq j^*$, owing to the definition that $\widetilde{A}_{ji} \leq A_{ji}$ for all $i \in [n], j \in [m]$. Defining $\delta^\flat = \frac{\mu\beta}{2A_{j^\flat,i}} \leq \delta$, the objective decrease is lower bounded as

$$\begin{aligned}
f_\mu(\mathsf{x}_k) - f_\mu(\mathsf{y}_k^{(i)}) &= \int_{\tau=0}^{\delta} \langle -\nabla f_\mu(\mathsf{x}_k + \tau\mathbf{e}_i), \mathbf{e}_i\rangle d\tau = \int_{\tau=0}^{\delta} \big(\langle A_{:i}, p(\mathsf{x}_k + \tau\mathbf{e}_i)\rangle - 1\big) d\tau \\
&\geq \int_{\tau=0}^{\delta^\flat} \big(\langle A_{:i}, p(\mathsf{x}_k + \tau\mathbf{e}_i)\rangle - 1\big) d\tau \\
&= \underbrace{\int_{\tau=0}^{\delta^\flat} \Big(-1 + \sum_{j\leq j^\flat} A_{j,i} \cdot p_j(\mathsf{x}_k + \tau\mathbf{e}_i)\Big) d\tau}_{I} + \underbrace{\sum_{j>j^\flat} \int_{\tau=0}^{\delta^\flat} A_{j,i} \cdot p_j(\mathsf{x}_k + \tau\mathbf{e}_i) d\tau}_{I'}
\end{aligned}$$

where the inequality is because $\delta^\flat \leq \delta$ and $\langle A_{:i}, p(\mathsf{x}_k + \tau\mathbf{e}_i)\rangle \geq 1$ for all $\tau \leq \delta$ (see the proof of part (a)).

**Part *I*.** To lower bound $I$, we use the monotonicity of $p_j(\cdot)$ and obtain that

$$I = \int_{\tau=0}^{\delta^\flat} \Big(-1 + \sum_{j\leq j^\flat} A_{j,i} \cdot p_j(\mathsf{x}_k + \tau\mathbf{e}_i)\Big) d\tau \geq \delta^\flat \cdot \Big(-1 + \sum_{j\leq j^\flat} A_{j,i} \cdot p_j(\mathsf{x}_k + \delta^\flat\mathbf{e}_i)\Big) \ .$$

However, our choice of $\delta^\flat = \frac{\mu\beta}{2A_{j^\flat,i}} \leq \frac{\mu\beta}{2A_{j,i}}$ for all $j \leq j^\flat$ ensures that

$$\sum_{j \leq j^\flat} A_{j,i} \cdot p_j(\mathsf{x}_k + \delta^\flat \mathbf{e}_i) \geq \sum_{j \leq j^\flat} A_{j,i} \cdot p_j(\mathsf{x}_k) \cdot e^{\frac{-A_{j,i} \cdot \delta^\flat}{\mu}} \geq \sum_{j \leq j^\flat} A_{j,i} \cdot p_j(\mathsf{x}_k) \cdot (1 - \beta/2) \ .$$

Therefore, we obtain that

$$I \geq \delta^\flat \Big( -1 + (1-\beta/2) \sum_{j \leq j^\flat} A_{j,i} \cdot p_j(\mathsf{x}_k) \Big) \geq \frac{\delta^\flat}{3} \Big( -1 + \sum_{j \leq j^\flat} A_{j,i} \cdot p_j(\mathsf{x}_k) \Big) \ ,$$

where the inequality is because $\big(\frac{2}{3} - \frac{\beta}{2}\big) \sum_{j \leq j^\flat} A_{j,i} \cdot p_j(\mathsf{x}_k) \geq \frac{4-3\beta}{6} \cdot (1+\beta) \geq \frac{2}{3}$ whenever $\beta \leq \frac{1}{3}$ (or equivalently, whenever $\varepsilon \leq 1/9$).

Now, suppose that $\sum_{j \leq j^\flat} \widetilde{A}_{j,i} \cdot p_j(\mathsf{x}_k) - (1+\beta) = b \cdot \widetilde{A}_{j^\flat,i} \cdot p_{j^\flat}(\mathsf{x}_k)$ for some $b \in [0,1]$. Note that we can do so by the very definition of $j^\flat$. Then, we must have

$$\begin{aligned} -1 + \sum_{j \leq j^\flat} A_{j,i} \cdot p_j(\mathsf{x}_k) &\geq -1 + \sum_{j < j^\flat} \widetilde{A}_{j,i} \cdot p_j(\mathsf{x}_k) + A_{j^\flat,i} \cdot p_{j^\flat}(\mathsf{x}_k) \\ &= -1 + (1+\beta) - (1-b)\widetilde{A}_{j^\flat,i} \cdot p_{j^\flat}(\mathsf{x}_k) + A_{j^\flat,i} \cdot p_{j^\flat}(\mathsf{x}_k) \\ &\geq \beta + b \cdot A_{j^\flat,i} \cdot p_{j^\flat}(\mathsf{x}_k) \ . \end{aligned}$$

Therefore, we conclude that

$$\begin{aligned} I &\geq \frac{\delta^\flat}{3} \Big( -1 + \sum_{j \leq j^\flat} A_{j,i} \cdot p_j(\mathsf{x}_k) \Big) > \frac{\delta^\flat}{3} \cdot b \cdot A_{j^\flat,i} \cdot p_{j^\flat}(\mathsf{x}_k) = \frac{\mu\beta}{6\widetilde{A}_{j^\flat,i}} \cdot b \cdot \widetilde{A}_{j^\flat,i} \cdot p_{j^\flat}(\mathsf{x}_k) \\ &= \frac{\mu\beta}{6\widetilde{A}_{j^\flat,i}} \cdot \Big( -(1+\beta) + \sum_{j \leq j^\flat} \widetilde{A}_{j,i} \cdot p_j(\mathsf{x}_k) \Big) \geq \frac{\mu\beta}{12} \cdot u_i^* \cdot \Big( -(1+\beta) + \sum_{j \leq j^\flat} \widetilde{A}_{j,i} \cdot p_j(\mathsf{x}_k) \Big) \ . \end{aligned}$$

Above, the last inequality is because $u_i^* \cdot \widetilde{A}_{j^\flat,i} \leq \langle \widetilde{A}_{j^\flat:}, u^* \rangle \leq 2$ by our definition of $\widetilde{A}$.

**Part $I'$.** To lower bound $I'$, consider every $j > j^\flat$ and the integral

$$\int_{\tau=0}^{\delta^\flat} A_{j,i} \cdot p_j(\mathsf{x}_k + \tau \mathbf{e}_i) d\tau \ .$$

Note that whenever $\tau \leq \frac{\mu\beta}{2A_{j,i}} \leq \frac{\mu\beta}{2A_{j^\flat,i}} = \delta^\flat$, we have that $p_j(\mathsf{x}_k + \tau\mathbf{e}_i) \geq p_j(\mathsf{x}_k) \cdot e^{-\beta/2} \geq \frac{1}{2} p_j(\mathsf{x}_k)$. Therefore,

$$\int_{\tau=0}^{\delta^\flat} A_{j,i} \cdot p_j(\mathsf{x}_k + \tau \mathbf{e}_i) d\tau \geq \int_{\tau=0}^{\frac{\mu\beta}{2A_{j,i}}} A_{j,i} \cdot p_j(\mathsf{x}_k + \tau \mathbf{e}_i) d\tau \geq \frac{\mu\beta}{2A_{j,i}} \cdot A_{j,i} \cdot \frac{1}{2} p_j(\mathsf{x}_k) \ .$$

This implies a lower bound on $I'$:

$$I' \geq \sum_{j > j^\flat} \frac{\mu\beta}{4A_{j,i}} \cdot A_{j,i} \cdot p_j(\mathsf{x}_k) \geq \frac{\mu\beta}{8} \cdot \sum_{j > j^\flat} u_i^* \cdot \widetilde{A}_{j,i} \cdot p_j(\mathsf{x}_k) \ ,$$

where again in the last inequality we have used $u_i^* \cdot \widetilde{A}_{j^\flat,i} \leq \langle \widetilde{A}_{j^\flat:}, u^* \rangle \leq 2$ by our definition of $\widetilde{A}$.

**Together.** Combining the lower bounds on $I$ and $I'$, we obtain

$$f_\mu(\mathsf{x}_k) - f_\mu(\mathsf{y}_k^{(i)}) \geq I + I' \geq \frac{\mu\beta}{12} \cdot u_i^* \cdot \Big( -(1+\beta) + \sum_{j=1}^{m} \widetilde{A}_{j,i} \cdot p_j(\mathsf{x}_k) \Big) = \frac{\mu\beta}{12} \cdot \langle -\widetilde{\eta}_k^{(i)}, u^* \rangle \ . \qquad \square$$

## E Proof of Lemma 3.3: Efficient Implementation of PacLPSolver

In this section, we illustrate how to implement each iteration of PacLPSolver to run in an expected $O(N/n)$ time. We maintain the following quantities

$$\mathsf{z}_k \in \mathbb{R}^n_{\geq 0}, \quad \mathsf{az}_k \in \mathbb{R}^m_{\geq 0}, \quad \mathsf{y}'_k \in \mathbb{R}^n, \quad \mathsf{ay}'_k \in \mathbb{R}^m, \quad B_{k,1}, B_{k,2} \in \mathbb{R}_+$$

throughout the algorithm, so as to ensure the following invariants are always satisfied

$$A\mathsf{z}_k = \mathsf{az}_k \ , \tag{E.1}$$

$$\mathsf{y}_k = B_{k,1} \cdot \mathsf{z}_k + B_{k,2} \cdot \mathsf{y}'_k \ , \qquad A\mathsf{y}'_k = \mathsf{ay}'_k \ . \tag{E.2}$$

It is clear that when $k = 0$, letting $\mathsf{az}_k = A\mathsf{z}_0$, $\mathsf{y}'_k = \mathsf{y}_0$, $\mathsf{ay}'_k = A\mathsf{y}_0$, $B_{k,1} = 0$, and $B_{k,2} = 1$, we can ensure that all the invariants are satisfied initially. We denote $\|A_{:i}\|_0$ the number of nonzeros elements in vector $A_{:i}$. In each iteration $k = 1, 2, \dots, T$:

- The step $\mathsf{x}_k = \tau \mathsf{z}_{k-1} + (1-\tau)\mathsf{y}_{k-1}$ does not need to be implemented.

- The value $\nabla_i f(\mathsf{x}_k)$ requires the knowledge of $p_j(\mathsf{x}_k) = e^{\frac{1}{\mu}((A\mathsf{x}_k)_j - 1)}$ for each $j$ such that $A_{ij} \neq 0$. Accordingly, for each $j$, we need to know the value

  $(A\mathsf{x}_k)_j = \tau(A\mathsf{z}_{k-1})_j + (1-\tau)(A\mathsf{y}_{k-1})_j = \big(\tau + (1-\tau)B_{k-1,1}\big)\mathsf{az}_{k-1,j} + (1-\tau)B_{k-1,2}\mathsf{ay}'_{k-1,j}$ .

  This can be computed in $O(1)$ time for each $j$, and $O(\|A_{:i}\|_0)$ time in total.

- Recall the step $\mathsf{z}_k \leftarrow \arg\min_{z \in \Delta_{\mathsf{box}}} \big\{\frac{1}{2}\|z - \mathsf{z}_{k-1}\|_A^2 + \langle n\alpha_k \xi_k^{(i)}, z\rangle\big\}$ can be written as $\mathsf{z}_k = \mathsf{z}_{k-1} + \delta \mathbf{e}_i$ for some $\delta \in \mathbb{R}$ that can be computed in $O(1)$ time (see Proposition 3.2). Observe also $\mathsf{z}_k = \mathsf{z}_{k-1} + \delta \mathbf{e}_i$ yields $\mathsf{y}_k = \tau \mathsf{z}_{k-1} + (1-\tau)\mathsf{y}_{k-1} + \frac{\delta \mathbf{e}_i}{n\alpha_k L}$ due to Line 6 and Line 10 of Algorithm 1. Therefore, we perform two explicit updates on $\mathsf{z}_k$ and $\mathsf{az}_k$ as

  $$\mathsf{z}_k \leftarrow \mathsf{z}_{k-1} + \delta \mathbf{e}_i \ , \quad \mathsf{az}_k \leftarrow \mathsf{az}_{k-1} + \delta A_{:i}$$

  and two implicit updates on $\mathsf{y}_k$ as

  $$B_{k,1} = \tau + (1-\tau)B_{k-1,1} \quad , \quad B_{k,2} = (1-\tau)B_{k-1,2} \ ,$$
  $$\mathsf{y}'_k \leftarrow \mathsf{y}'_{k-1} + \delta \mathbf{e}_i \cdot \Big(-\frac{B_{k,1}}{B_{k,2}} + \frac{1}{n\alpha_k L}\frac{1}{B_{k,2}}\Big) \quad , \quad \mathsf{ay}'_k \leftarrow \mathsf{ay}'_{k-1} + \delta A_{:i} \cdot \Big(-\frac{B_{k,1}}{B_{k,2}} + \frac{1}{n\alpha_k L}\frac{1}{B_{k,2}}\Big)$$

  It is not hard to verify that after these updates, $A\mathsf{y}'_k = \mathsf{ay}'_k$ and we have

  $$\begin{aligned}
  B_{k,1} \cdot \mathsf{z}_k + B_{k,2} \cdot \mathsf{y}'_k &= B_{k,1} \cdot \big(\mathsf{z}_{k-1} + \delta \mathbf{e}_i\big) + B_{k,2} \cdot \Big(\mathsf{y}'_{k-1} + \delta \mathbf{e}_i \cdot \Big(-\frac{B_{k,1}}{B_{k,2}} + \frac{1}{n\alpha_k L}\frac{1}{B_{k,2}}\Big)\Big) \\
  &= B_{k,1} \cdot \mathsf{z}_{k-1} + B_{k,2} \cdot \Big(\mathsf{y}'_{k-1} + \delta \mathbf{e}_i \cdot \Big(\frac{1}{n\alpha_k L}\frac{1}{B_{k,2}}\Big)\Big) \\
  &= B_{k,1} \cdot \mathsf{z}_{k-1} + B_{k,2} \cdot \mathsf{y}'_{k-1} + \frac{\delta \mathbf{e}_i}{n\alpha_k L} \\
  &= \big(\tau + (1-\tau)B_{k-1,1}\big) \cdot \mathsf{z}_{k-1} + \big((1-\tau)B_{k-1,2}\big) \cdot \mathsf{y}'_{k-1} + \frac{\delta \mathbf{e}_i}{n\alpha_k L} \\
  &= \tau \mathsf{z}_{k-1} + (1-\tau)\mathsf{y}_{k-1} + \frac{\delta \mathbf{e}_i}{n\alpha_k L} = \mathsf{y}_k \ ,
  \end{aligned}$$

  so the invariant $\mathsf{y}_k = B_{k,1} \cdot \mathsf{z}_k + B_{k,2} \cdot \mathsf{y}'_k$ also holds. In sum, after performing updates on $A\mathsf{z}_k$ and $\mathsf{ay}'_k$ in time $O(\|A_{:i}\|_0)$, we can ensure that the invariants in (E.1) and (E.2) are satisfied at iteration $k$.

In sum, we only need $O(\|A_{:i}\|_0)$ time to perform the updates in PacLPSolver for an iteration $k$ if the coordinate $i$ is selected. Therefore, each iteration of PacLPSolver can be implemented to run in an expected $O(\mathbf{E}_i[\|A_{:i}\|_0]) = O(N/n)$ time.

## F Proof of Lemma 6.5: Efficient Implementation of `CovLPSolver`

In this section we illustrate how to implement each iteration of `CovLPSolver` to run in an expected $O(N/n)$ time. We maintain the following quantities

$$\mathsf{z}'_k \in \mathbb{R}^n_+, \quad \mathsf{sz}_k \in \mathbb{R}_+, \quad \mathsf{sumz}_k \in \mathbb{R}_+, \quad \mathsf{az}'_k \in \mathbb{R}^m_{\geq 0}, \quad \mathsf{y}'_k \in \mathbb{R}^n, \quad \mathsf{ay}'_k \in \mathbb{R}^m, \quad B_{k,1}, B_{k,2} \in \mathbb{R}_+$$

throughout the algorithm, so as to maintain the following invariants

$$\mathsf{z}_k = \mathsf{z}'_k/\mathsf{sz}_k, \qquad \mathsf{sumz}_k = \mathbb{1}^T \mathsf{z}'_k, \qquad A\mathsf{z}_k = \mathsf{az}'_k/\mathsf{sz}_k, \tag{F.1}$$

$$\mathsf{y}_k = B_{k,1} \cdot \mathsf{z}'_k + B_{k,2} \cdot \mathsf{y}'_k, \qquad A\mathsf{y}_k = \mathsf{ay}'_k \ . \tag{F.2}$$

It is clear that when $k = 0$, letting $\mathsf{z}'_k = \mathsf{z}_0$, $\mathsf{sz}_k = 1$, $\mathsf{sumz}_k = \mathbb{1}^T \mathsf{z}_0$, $\mathsf{az}'_k = A\mathsf{z}_0$, $\mathsf{y}'_k = \mathsf{y}_0$, $\mathsf{ay}'_k = A\mathsf{y}_0$, $B_{k,1} = 0$, and $B_{k,2} = 1$, we can ensure that all the invariants are satisfied initially.

We denote by $\|A_{:i}\|_0$ the number of nonzero elements in vector $A_{:i}$. In each iteration $k = 1, 2, \ldots, T$:

- The step $\mathsf{x}_k = \tau \mathsf{z}_{k-1} + (1-\tau)\mathsf{y}_{k-1}$ does not need to be implemented.
- The value $p_j(\mathsf{x}_k) = e^{\frac{1}{\mu}(1-(A\mathsf{x}_k)_j)}$ for each $j$ only requires the knowledge of
$$(A\mathsf{x}_k)_j = \tau (A\mathsf{z}_{k-1})_j + (1-\tau)(A\mathsf{y}_{k-1})_j = \big(\tau + (1-\tau)B_{k-1,1}\big)\frac{\mathsf{az}'_{k-1,j}}{\mathsf{sz}_{k-1}} + (1-\tau)B_{k-1,2}\mathsf{ay}'_{k-1,j} \ .$$
This can be computed in $O(1)$ time.
- The value $\nabla_i f(\mathsf{x}_k)$ requires the knowledge of $p_j(\mathsf{x}_k)$ for each $j \in [m]$ such that $A_{ij} \neq 0$. Since we have $\|A_{:i}\|_0$ such $j$'s, we can compute $\nabla_i f(\mathsf{x}_k)$ in $O(\|A_{:i}\|_0)$ time.
- Letting $\delta = (1+\gamma)n\alpha_k \xi^{(i)}_{k,i}$, recall that the mirror step $\mathsf{z}_k \leftarrow \arg\min_{z \in \Delta_{\mathsf{simplex}}} \big\{V_{\mathsf{z}_{k-1}}(z) + \langle \delta \mathbf{e}_i, z\rangle\big\}$ has a very simple form (see Proposition 6.4): first multiply the $i$-th coordinate of $\mathsf{z}_{k-1}$ by $e^{-\delta}$ and then, if the sum of all coordinates have exceeded $2\mathsf{OPT}'$, scale everything down so as to sum up to $2\mathsf{OPT}'$. This can be implemented as follows: setting $\delta_1 = \mathsf{z}'_{k-1,i}(e^{-\delta} - 1)$,
$$\mathsf{z}'_k \leftarrow \mathsf{z}'_{k-1} + \delta_1 \mathbf{e}_i \quad , \quad \mathsf{az}'_k \leftarrow \mathsf{az}'_{k-1} + \delta_1 A_{:i} \ ,$$
$$\mathsf{sumz}_k \leftarrow \mathsf{sumz}_{k-1} + \delta_1 \quad , \quad \mathsf{sz}_k \leftarrow \mathsf{sz}_k \cdot \max\Big\{1, \tfrac{\mathsf{sumz}_k}{\mathsf{sz}_{k-1} \cdot 2\mathsf{OPT}'}\Big\} \ .$$
These updates can be implemented to run in $O(\|A_{:i}\|_0)$ time, and they together ensure that the invariants in (F.1) are satisfied at iteration $k$.
- Recall that the gradient step is of the form $\mathsf{y}_k \leftarrow \mathsf{x}_k + \delta_2 \cdot \mathbf{e}_i$ for some value $\delta_2 \geq 0$. This value $\delta_2$ can be computed in $O(\|A_{:i}\|_0)$ time, since each $p_j(\mathsf{x}_k)$ can be computed in $O(1)$ time, and we can sort the rows of each column of $A$ by preprocessing.

  Since $\mathsf{y}_k = \mathsf{x}_k + \delta_2 \cdot \mathbf{e}_i = \tau \mathsf{z}_{k-1} + (1-\tau)\mathsf{y}_{k-1} + \delta_2 \mathbf{e}_i$, we can implement this update by letting
$$B_{k,1} = \tfrac{\tau}{\mathsf{sz}_{k-1}} + (1-\tau)B_{k-1,1} \quad , \quad B_{k,2} = (1-\tau)B_{k-1,2}$$
$$\mathsf{y}'_k \leftarrow \mathsf{y}'_{k-1} + \mathbf{e}_i \cdot \Big(-\tfrac{B_{k,1}\delta_1}{B_{k,2}} + \tfrac{\delta_2}{B_{k,2}}\Big) \quad , \quad \mathsf{ay}'_k \leftarrow \mathsf{ay}'_{k-1} + A_{:i} \cdot \Big(-\tfrac{B_{k,1}\delta_1}{B_{k,2}} + \tfrac{\delta_2}{B_{k,2}}\Big)$$
  It is not hard to verify that after these updates, $\mathsf{ay}'_k = A\mathsf{y}'_k$ and we have
$$B_{k,1} \cdot \mathsf{z}'_k + B_{k,2} \cdot \mathsf{y}'_k = B_{k,1} \cdot \big(\mathsf{z}'_{k-1} + \delta_1 \mathbf{e}_i\big) + B_{k,2} \cdot \Big(\mathsf{y}'_{k-1} + \mathbf{e}_i \cdot \Big(-\frac{B_{k,1}\delta_1}{B_{k,2}} + \frac{\delta_2}{B_{k,2}}\Big)\Big)$$
$$= B_{k,1} \cdot \mathsf{z}'_{k-1} + B_{k,2} \cdot \big(\mathsf{y}'_{k-1} + \delta_2 \mathbf{e}_i / B_{k,2}\big)$$
$$= B_{k,1} \cdot \mathsf{z}'_{k-1} + B_{k,2} \cdot \mathsf{y}'_{k-1} + \delta_2 \mathbf{e}_i$$
$$= \big(\frac{\tau}{\mathsf{sz}_{k-1}} + (1-\tau)B_{k-1,1}\big) \cdot \mathsf{z}'_{k-1} + \big((1-\tau)B_{k-1,2}\big) \cdot \mathsf{y}'_{k-1} + \delta_2 \mathbf{e}_i$$

$= \tau \mathsf{z}_{k-1} + (1-\tau)\mathsf{y}_{k-1} + \delta_2 \mathbf{e}_i = \mathsf{y}_k$ ,

so that the invariant $\mathsf{y}_k = B_{k,1} \cdot \mathsf{z}'_k + B_{k,2} \cdot \mathsf{y}'_k$ is also satisfied. In sum, after running time $O(\|A_{:i}\|_0)$, we can ensure that the invariants in (F.2) are satisfied at iteration $k$.

In sum, we only need $O(\|A_{:i}\|_0)$ time to perform the updates in `CovLPSolver` for an iteration $k$ if the coordinate $i$ is selected. Therefore, each iteration of `CovLPSolver` can be implemented to run in an expected $O(\mathbf{E}_i[\|A_{:i}\|_0]) = O(N/n)$ time.